\documentclass[prl,twocolumn,superscriptaddress]{revtex4-1}

\usepackage{color}
\usepackage[english]{babel}
\usepackage[T1]{fontenc}
\usepackage[latin9]{inputenc}
\usepackage{latexsym}
\usepackage{float}
\usepackage{amsmath}
\usepackage{graphicx}
\usepackage{times}   
\usepackage{esint}
\usepackage{verbatim}
\usepackage{subfigure}
\usepackage{color}
\usepackage{amsmath, amssymb, mathrsfs}
\usepackage[unicode=true,pdfusetitle, bookmarks=false,colorlinks=true,citecolor=blue,urlcolor=blue,linkcolor=red]{hyperref}

\makeatletter

\newcommand{\beq}{\begin{eqnarray}}
\newcommand{\eeq}{\end{eqnarray}}
\newcommand{\be}{\begin{equation}}
\newcommand{\ee}{\end{equation}}
\newcommand{\prlsec}[1]{\emph{#1---}}

\begin{document}

\title{Phase diagram of the $Z_3$ Parafermionic Chain with Chiral Interactions}
\author{Ye Zhuang, Hitesh J. Changlani, Norm M. Tubman, Taylor L. Hughes}
\affiliation{Department of Physics and Institute of Condensed Matter Theory, University of Illinois, Urbana, Illinois 61801, USA\\}
\date{\today}

\begin{abstract}

Parafermions are exotic quasiparticles with non-Abelian fractional statistics that can be realized and stabilized
in 1-dimensional models that are generalizations of the Kitaev p-wave wire. We study the simplest generalization, i.e. the $Z_3$ parafermionic chain. Using a Jordan-Wigner transform we focus on the equivalent three-state chiral clock model, and study its rich phase diagram using the density matrix renormalization group technique. 
We perform our analyses using quantum entanglement diagnostics which allow us to determine phase boundaries, and the nature of the phase transitions. In particular, we study the transition between the topological and trivial phases, as well as to an intervening incommensurate phase which appears in a wide region of the phase diagram. The phase diagram is predicted to contain a Lifshitz type transition which we confirm using entanglement measures.  We also attempt to locate and characterize a putative tricritical point in the phase diagram where the three above mentioned phases meet at a single point. 

\end{abstract}

\maketitle

\newpage


\prlsec{Introduction}
There has been concerted effort to engineer systems with stable Majorana bound states, 
and other anyonic quasiparticles, for use in the topological quantum computation architecture~\cite{nayak2008rev,fu2008,fu2009,lutchyn2010,oreg2010,alicea2011non,alicea2012}.
For example, there has been recent progress in attempts to 
isolate Majorana bound states in quantum nanowires~\cite{oreg2010,mourik2012signatures, rokhinson2012fractional,finck2013} 
and in superconductor surfaces implanted with a line of magnetic impurities~\cite{nadj2014observation}. 
These quasi-1D systems effectively realize a version of the Kitaev p-wave wire model~\cite{kitaev2001}, 
and are predicted to have a gapped topological phase which supports characteristic Majorana bound states at the ends of the wire.

While the boundary modes in these heterostructure systems are non-Abelian anyons, they are unfortunately known to be
insufficient for universal quantum computation. 
A possible remedy for this problem has been to look for more exotic non-Abelian excitations.
For example, Fendley has recently suggested exploring one-dimensional $Z_N$ 
para-fermionic models which support  topological phases with more computationally efficient non-Abelian anyon bound states~\cite{fendley2012}.
Still, the $Z_N$ non-Abelian anyons are not able to perform universal quantum computation, however they can be leveraged to create a 2D phase with Fibonaccci anyons, which are universal~\cite{mong2014univerdal}.
These promising features have spurred wide spread interest
in these models, and has led to many analytical and numerical 
studies, including several experimental proposals for realizing these topological phases
~\cite{2010ortiz,Lindner2012, cheng2012, clarke2013, vaezi2013, rapp2013, hastings2013, burrello2013, bondesan2013, barkeshli2013, barkeshli2014experimental, cobanera2014, teo2014luttinger, oreg2014, orth2014non, klinovaja2014parafermions, zhang2014, li2014, tsvelik2014, klinovaja2014time, klinovaja2014kramers, mong2014parafermionic, alicea2014, milsted2014,2014rigol}.

In this work, we continue along these lines of research by exploring
the rich phase diagram of the $Z_3$ para-fermionic chain; though for ease of calculation we actually study the Jordan-Wigner transformed para-fermionic chain\cite{fradkin1980}, including chiral interactions. The resulting model is the three state \emph{chiral} clock model. This model re-surfaced in this context in Ref.~\onlinecite{fendley2012} as a candidate for exhibiting non-Abelian bound states beyond Majorana fermions. 
It was shown analytically that  para-fermionic boundary zero modes 
can exist in this model  when spatial-parity and time-reversal symmetries are broken 
via chiral interactions~\cite{fendley2012}. This  was
 verified numerically in Ref.~\onlinecite{jermyn2014}, which confirms that chiral interactions can help to stabilize the boundary zero modes, although the zero modes themselves are more fragile than one might initially expect.  

Here we are interested in studying the full phase diagram of the chiral clock model as a function of two chiral-interaction phase-parameters $(\theta,\phi),$  as well as the relative strength of the nearest neighbor coupling $(J)$ to the local Zeeman field $(f)$. Using entanglement techniques, we have been able to locate the phase boundaries that 
separate the topological phase from the trivial gapped phase, and a critical incommensurate phase, 
the latter of which has no analog in the Kitaev p-wave wire model. We have conclusively identified the region 
in which there is a topological phase, and have explored the nature of the quantum phase transitions in and out of 
the three adjoining phases. 
In addition, by studying oscillatory properties of the system in, or near, the incommensurate phase, we 
establish the approximate location of a putative tricritical point\cite{ostlund1981,howes1983}, and further support the entanglement signatures that were recently proposed for identifying Lifshitz transitions\cite{rodney2013}. 

The article is arranged as follows. We first discuss the details of the model, 
and the criteria used to map out its phase diagram.  For our numerical simulations, 
the density matrix renormalization group (DMRG)~\cite{white1992, schollwock2011} algorithm 
is employed, as it gives immediate access to the entanglement entropy (EE), and 
therefore the central charge, at putative critical points/regions in the phase diagram~\cite{calabrese2004}.
Next, we discuss the general features of the phase diagram and locate regions 
in the topological phase (where para-fermion boundary modes may exist). We also discuss the nature of the phase transitions out of the topological phase.
For part of our study we discuss our observations pertaining to a critical incommensurate phase, and the possibility of a 
tricritical point~\cite{ostlund1981,howes1983} in the phase diagram at the intersection of the topological, trivial, and incommensurate phases. 
We also find a region of the phase diagram which exhibits the critical entanglement features of a Lifshitz transition~\cite{rodney2013}.
Finally, we conclude by discussing future directions and possible 
relevance to experiments looking for para-fermions. We also include four appendices which discuss some subtleties of the numerical analysis. 


\prlsec{The Model}
For our study we use the 1D 3-state ($Z_3$) chiral 
clock model~\cite{ostlund1981,huse1981,fendley2012,jiang2012identifying}. 
The Hamiltonian for the $3$-state chiral clock model is: 
\begin{equation}
H_3 = -f \sum_{j=1}^L \tau_j^{\dagger} e^{-i\phi}  - J \sum_{j=1}^{L-1} \sigma_j^{\dagger} \sigma_{j+1} e^{-i\theta} + \textrm{h.c.}
\label{eq:H3}
\end{equation}
following the notation in previous work~\cite{fendley2012}, where 
$f$, $J$, $\theta$ and $\phi$ are scalar parameters, 
and $\sigma_i$ and $\tau_i$ are local three state spin operators on site $i$. 
The spin operators have the properties $\tau^3=\sigma^3=I$, 
$\sigma\tau=\omega\;\tau\sigma$, where $\omega=e^{2\pi i/3}$.
Specifically, we use the matrix representation
\begin{displaymath}
\tau = \left(
\begin{array}{ccc}
1 & 0 & 0 \\
0 & \omega & 0 \\
0 & 0 & \omega^2 \\
\end{array}
\right),\quad
\sigma = \left(
\begin{array}{ccc}
0 & 1 & 0 \\
0 & 0 & 1 \\
1 & 0 & 0 \\
\end{array}
\right).
\end{displaymath}

The chiral clock model is related to the para-fermionic chain through a Jordan-Wigner transformation~\cite{fradkin1980,fendley2012}, 
similar to the well-known, analogous case that the Kitaev p-wave wire is related to the transverse-field Ising model 
via the same type of transformation. The parafermion operators are defined as
\begin{equation}
\chi_j = \left( \prod_{k=1}^{j-1} \tau_k \right) \sigma_j,
\end{equation}
\begin{equation}
\psi_j = \left( \prod_{k=1}^{j-1} \tau_k \right) \sigma_j \tau_j,
\end{equation}
at site $j$. The corresponding para-fermionic Hamiltonian is
\begin{equation}
H_3 = -f\sum_{j=1}^L  \psi_j^{\dagger} \chi_j e^{-i\phi} - J \omega^2 \sum_{j=1}^{L-1} \psi^{\dagger}_j \chi_{j+1} e^{-i\theta} + \textrm{h.c.}
\end{equation}

The chiral clock model has a global $Z_3$ symmetry 
that can be represented with $\chi \equiv \prod_{j=1}^L \tau_j^{\dagger}  \equiv e^{\frac{2\pi i}{3}\mathcal{Z}}.$ 
Here $\mathcal{Z}$ is the generator of the symmetry, and has three different eigenvalues 0,1,2.
In addition, when all of the coefficients of the Hamiltonian are real, i.e. when the system is a $Z_3$-ferromagnet or 
$Z_3$-anti-ferromagnet Hamiltonian, then the Hamiltonian is invariant under time-reversal, charge-conjugation, and 
parity symmetries. 
This can be easily seen from the following definitions of these symmetries.
Charge conjugation $C$ acts on the spin operators via 
$C \sigma_j C = \omega^2\sigma_j^{\dagger}$, $C \tau_j C = \tau_j^{\dagger}$, $ C^2 = 1$.
As an aside, note that charge conjugation, together with the $Z_3$ symmetry, forms the $S_3$ permutation symmetry, i.e. the symmetry obeyed when the $3$-state clock model is restricted to the $3$-state Potts model.
Time reversal $T$ acts on the spin operators via 
$T \sigma_j T = \sigma_j$, $T \tau_j T = \tau_j^{\dagger}$, $T^2 = 1,$ and complex conjugates any scalar coefficients.
Spatial parity $P$ acts on the spin operators via 
$P \sigma_j P = \sigma_{-j}$, $P \tau_j P = \tau_{-j}$, $P^2 = 1$.
Finally, we note two things: (i) due to the symmetry of the Hamiltonian with respect to $\phi$ and $\theta$, we only need to 
consider the region of the phase diagram where $\phi$ and $\theta$ each range from 0 to $\frac{\pi}{3}$, 
and (ii) for $f=J$, the system is self-dual along the line $\phi=\theta$. 
The details of these two properties are in Appendix~\ref{app:duality}.

There are many previously known results about 
this model (Eq. \ref{eq:H3}), beginning with the original proposals of
Ostlund~\cite{ostlund1981} and Huse~\cite{huse1981}. 
For example, the corresponding two-dimensional classical Hamiltonian for
$\phi=0$ was studied in Ref.~\onlinecite{ostlund1981}, and the
the one-dimensional quantum Hamiltonian 
was studied in Ref.~\onlinecite{howes1983} for the restricted case $\phi=\theta.$
One of the most important early results is that Eq.~\eqref{eq:H3} has  
a second order quantum phase transition 
at $f=J$ when  $\theta=\phi=0.$ 
At this point the model realizes the full $S_3$ permutation symmetry (instead of just $Z_3$), and the critical point 
is described by the critical conformal field theory for the $3$-state Potts model, which has  
central charge $4/5$ ~\cite{CFT}. In addition, the line 
$f\cos(3\phi)=J\cos(3\theta)$~\cite{Auyang} is known to be integrable 
and $\phi=\theta=\frac{\pi}{6}$ is super integrable~\cite{Albertini1989,McCoy1990}. Despite this, 
the knowledge of the location of some important critical 
points and their associated properties is an open question.

Generically, it is known that the phase diagram is divided up into two gapped regions, 
one of which is identified with small values of $f$ (compared with $J$), and the other with large values of $f.$ These regions are separated by continuous quantum phase transitions that we will identify and discuss further below. Using a more modern terminology, the gapped phase for small $f$ is a symmetry broken phase of the $3$-state clock model and it exactly corresponds to the  ``topological" phase in the Jordan-Wigner transformed para-fermionic chain. The gapped phase for large $f$ is a disordered phase of the $3$-state clock model, and maps onto the ``trivial" phase of the para-fermionic chain. This gives another example of a case where the degeneracy associated to symmetry breaking is mapped to topological degeneracy via the Jordan-Wigner transformation~\cite{chen2007, Greiter2014}. Hence, in either representation this phase 
has a three-fold ground-state degeneracy, which 
can be detected by measuring the ground state EE. 
On the other hand, the trivial phase is equivalent to the spin 
disordered phase, which does not have a generic ground-state degeneracy. 
The parameter $f$ is thus an important tuning parameter for the phase diagram, 
and analogous to the external transverse field in the Ising model. 

While we expect these general features to pervade the phase diagram, 
the phase space for generic $\theta$ and $\phi$ is largely unexplored. 
Additionally, it is known that the combination of the $Z_3,$ symmetry and the chiral nature of the 
interactions, gives rise to interesting behavior that cannot be found in the Majorana/Ising case. For example, this model supports a 
so-called ``incommensurate phase" which is 
not present in the transverse-field Ising model 
with chiral interactions~\cite{ostlund1981}.

This motivates the main objective of our article, which is to characterize the phases and the
nature of the phase transitions over the entire phase space.
We will show that there are two types of phase transitions 
that occur to destabilize the topological phase, and there is a large region 
of critical incommensurate phase that separates the topological from the trivial phase 
over a wide range of parameters. Let us now move on to a discussion of the methods we employ.


\prlsec{Methods}
We primarily use the spatial EE in order to characterize the phase diagram. 
This measure has been widely used to detect topological order in 2D~\cite{levin2006,kitaev2006topological}, 
and has been applied more recently to 1D topological phases~\cite{pollmann2010entanglement}. 
The EE can be derived by partitioning the system into two regions $A$ and $B,$ and then calculating the reduced density matrix 
of region $A$ by tracing over all the degrees of freedom in region $B.$
Mathematically, the reduced density matrix is given by $\rho_A \equiv \textrm{Tr}_B \rho $, and the 
corresponding entanglement entropy is defined to be:
\begin{equation}
S \equiv - \textrm{Tr} \left( \rho_A \ln \rho_A \right).
\end{equation}

There are two useful entanglement indicators we will employ to identify the phases and phase transitions for the chiral clock model.  
First, for the gapped regions of the phase digram, it is known that for one dimensional gapped systems the entanglement entropy 
increases with the the block size $l$ (the size of region $A$), and saturates when $l$ reaches the correlation length~\cite{calabrese2004}. 
Furthermore, if there is topological ground-state degeneracy we would expect an entanglement of order $\sim \log D$ 
where $D$ is the degeneracy~\cite{pollmann2010entanglement}.
To eliminate the most harmful finite-size effects we will take the central-cut, i.e., cutting the chain in half, to identify the nature of the gapped phases.

For critical regions of the phase diagram, it is known that the entanglement entropy will grow logarithmically with system size, and
the scaling is characterized by the central charge~\cite{calabrese2004}.
More specifically, for critical systems with open boundary conditions, 
the form of the entanglement scaling law is~\cite{calabrese2004}:
\begin{equation}
S = \frac{c}{6} \ln \left( \frac{2L}{\pi} \sin \frac{\pi l}{L} \right) + S_0
\label{eq:EE}
\end{equation}
where $l$ is the length of the subsystem, $c$ is the central charge, and $S_{0}$ contains the sub-leading corrections. Once we know the central charge we will have an important piece of information about the phase transition/critical phase, and can then appeal to previously known analytic results in restricted parts of the phase diagram to help further specify the phase diagram. Below we will see the efficacy of these two indicators for determining the phase diagram.

To arrive at  our results for the phase diagram (and to obtain reasonable estimates of the phase boundaries 
in the thermodynamic limit), we simulated Hamiltonians using open-boundary DMRG with 100 sites, and a bond dimension 
$m=100$. We find this to be sufficient for the phases with low entanglement entropy. For the critical phases, 
additional checks were performed with bond dimension $m=200.$
For establishing characteristics of other phases, for example, the region of critical incommensurate phase, larger lengths of 
$400$ sites were also tested.


\begin{figure}[htpb]
\centering
\includegraphics[width=\linewidth]{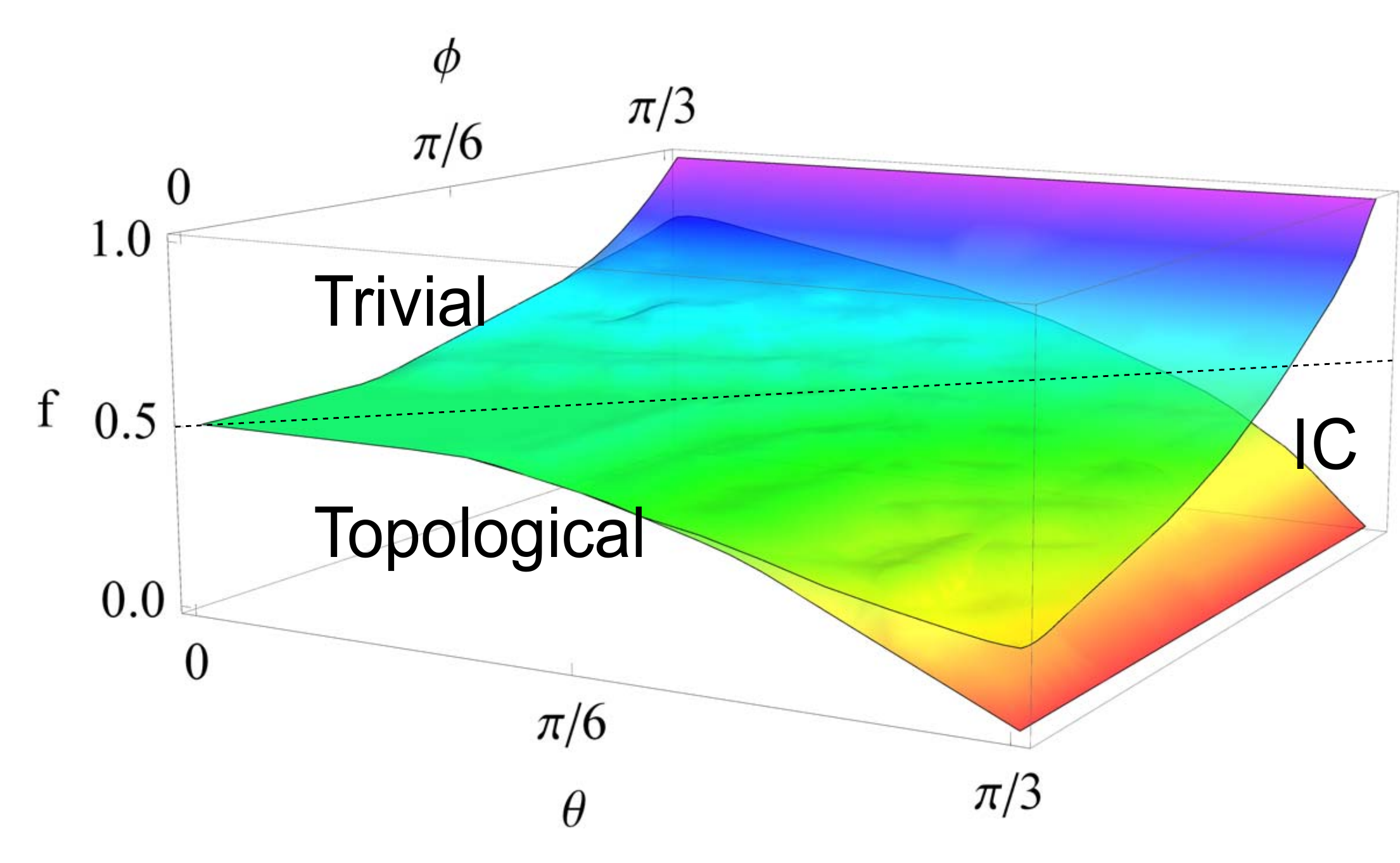} 
\caption{ Three-dimensional phase diagram of the chiral $3$-state clock model 
in terms of $f$, $\theta$ and $\phi$ with $J=1-f$. 
For details of the Hamiltonian see Eq.~\eqref{eq:H3}.
The topological, trivial, and incommensurate (IC) phases are indicated. 
The coloring is a function of the value of $f$ at the critical surface separating the 
phases. The dashed line that connects points (0, 0, 0.5) and ($\pi/3$, $\pi/3$, 0.5) is the self-dual line.
}\label{phase_3d}
\end{figure}

\begin{figure*}[htpb]
\subfigure[]{\includegraphics[width=0.32\linewidth]{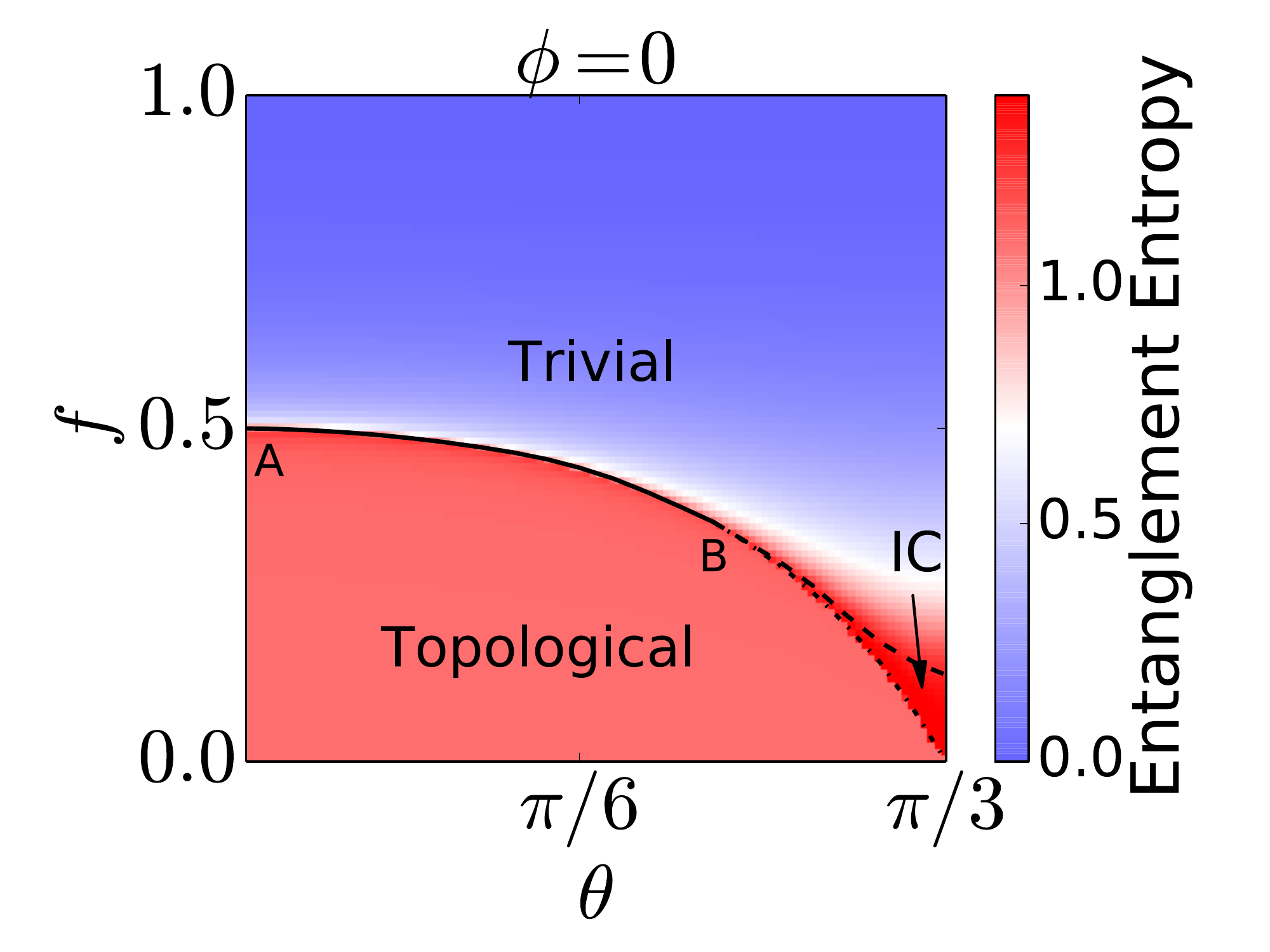} \label{phase_0}}
\subfigure[]{\includegraphics[width=0.32\linewidth]{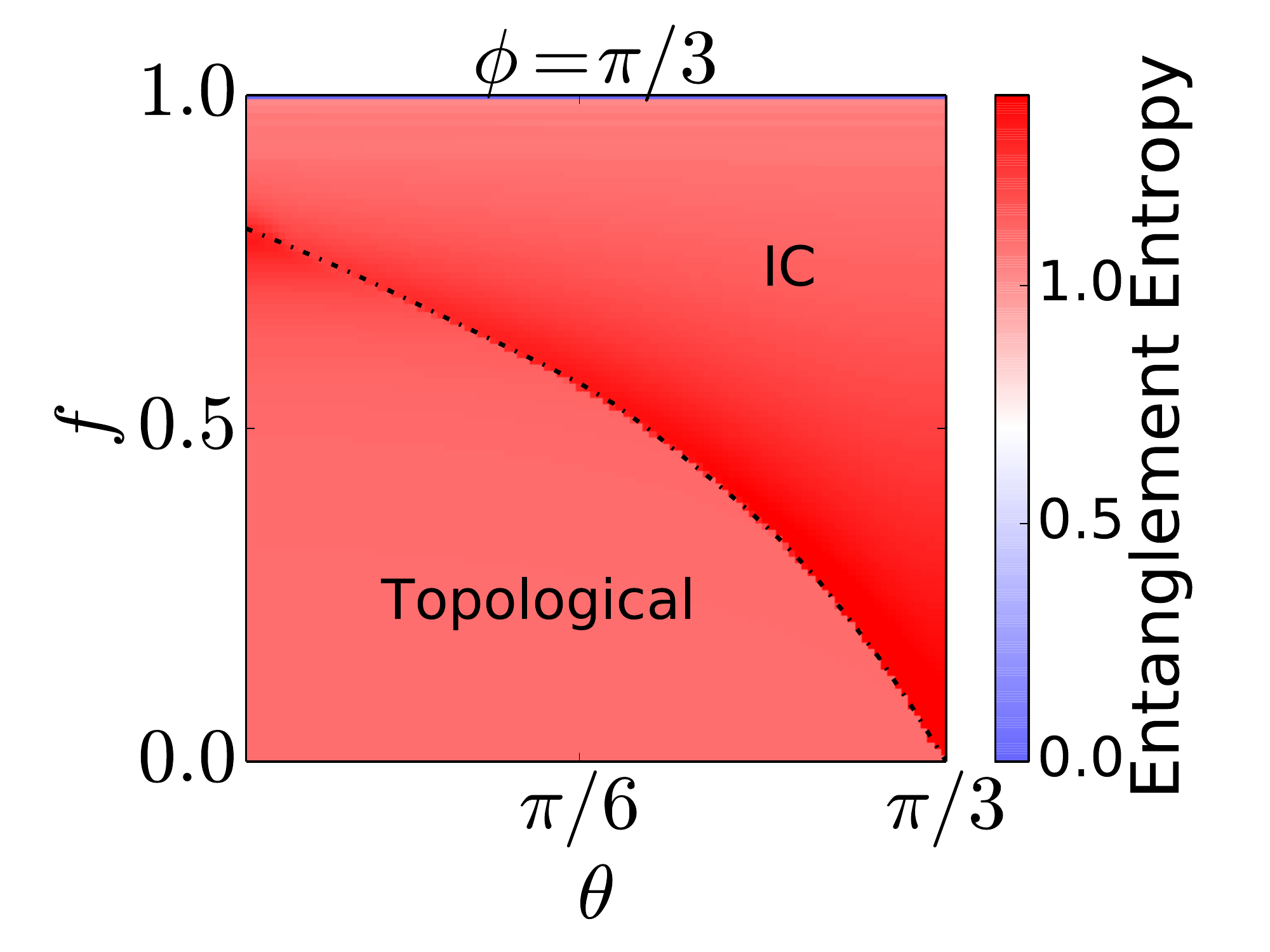} \label{phase_2}}
\subfigure[]{\includegraphics[width=0.32\linewidth]{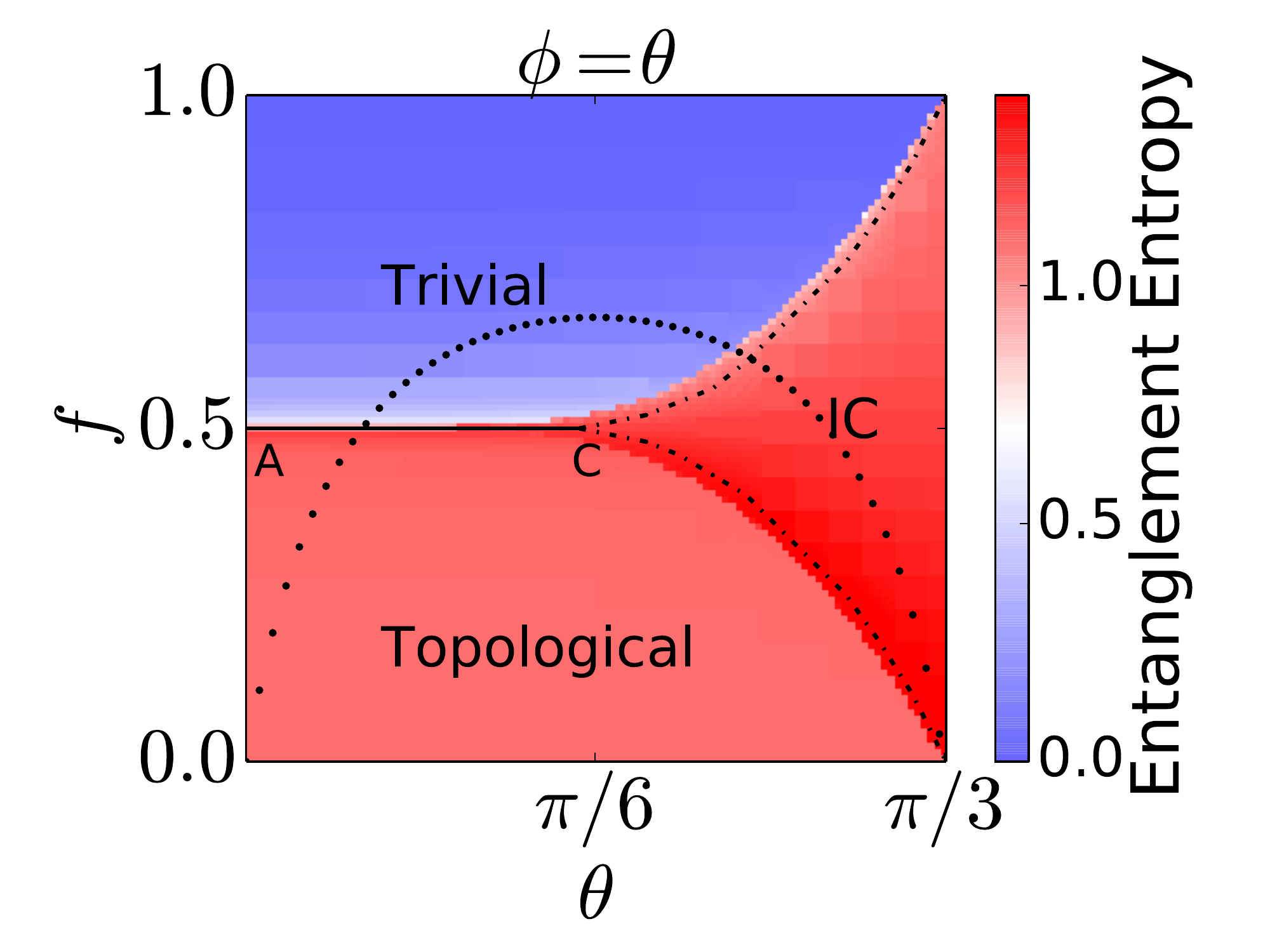} \label{phase_6}}
\subfigure[]{\includegraphics[width=0.32\linewidth]{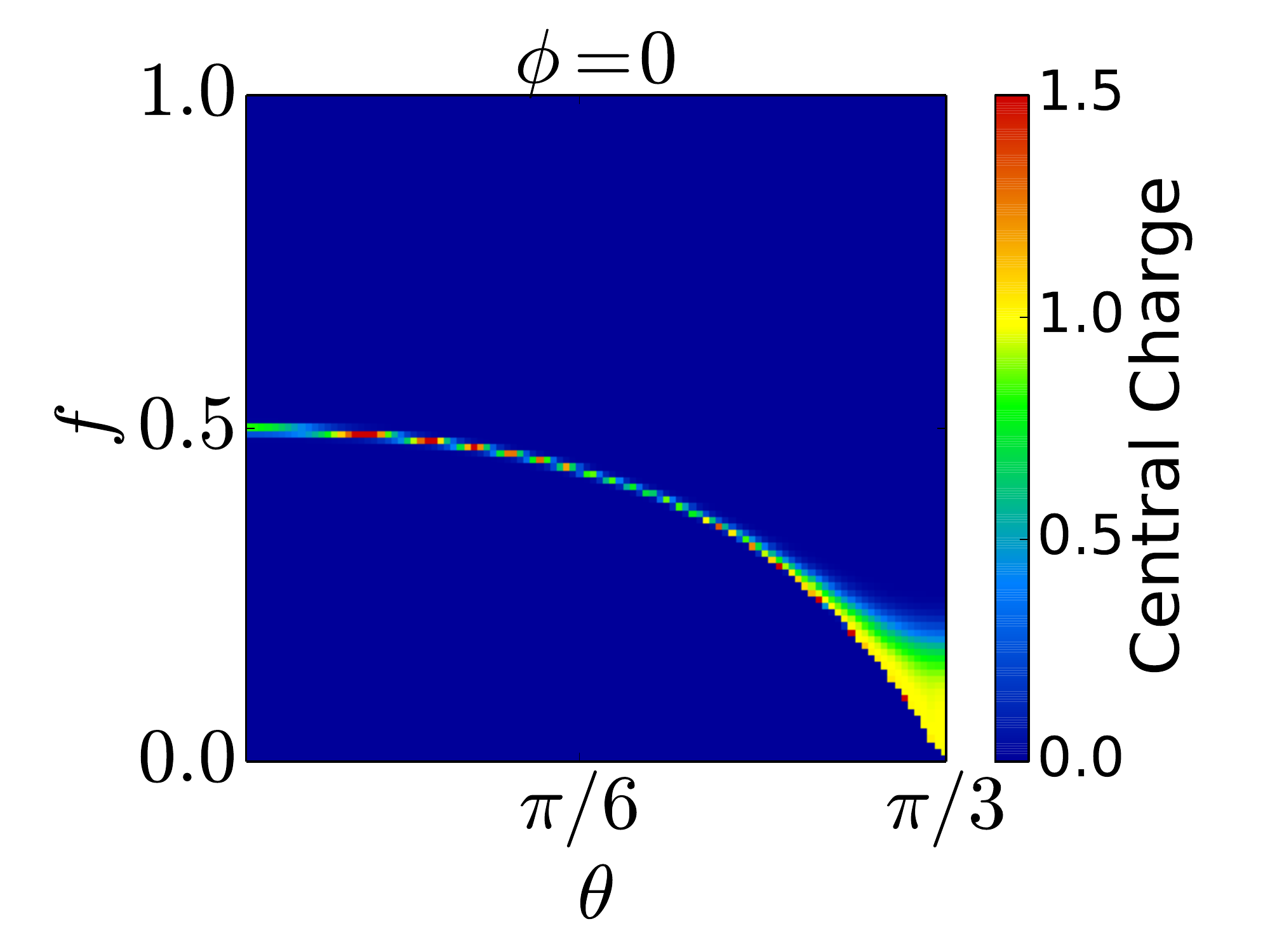} \label{phase_1}}
\subfigure[]{\includegraphics[width=0.32\linewidth]{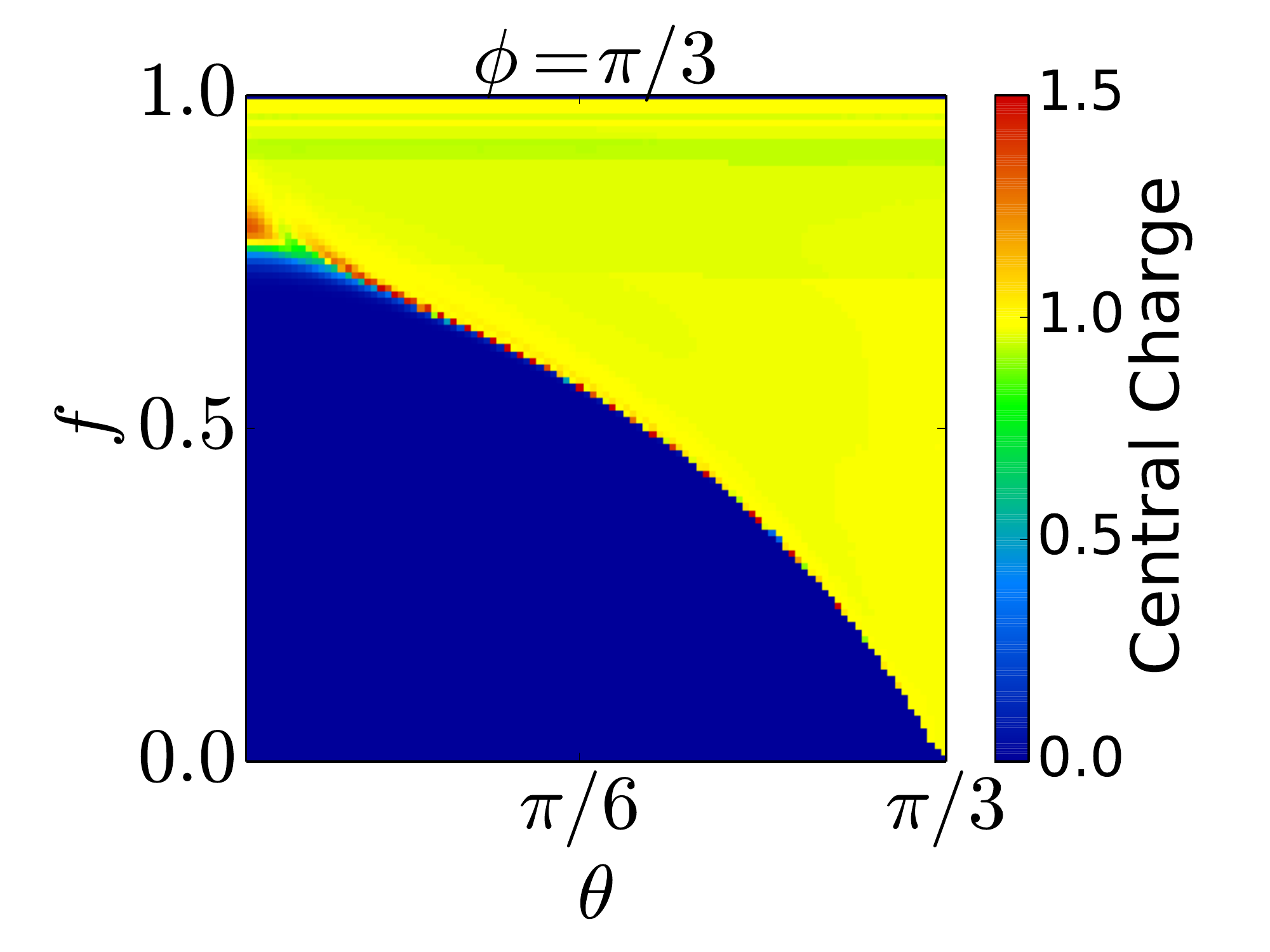} \label{phase_3}}
\subfigure[]{\includegraphics[width=0.32\linewidth]{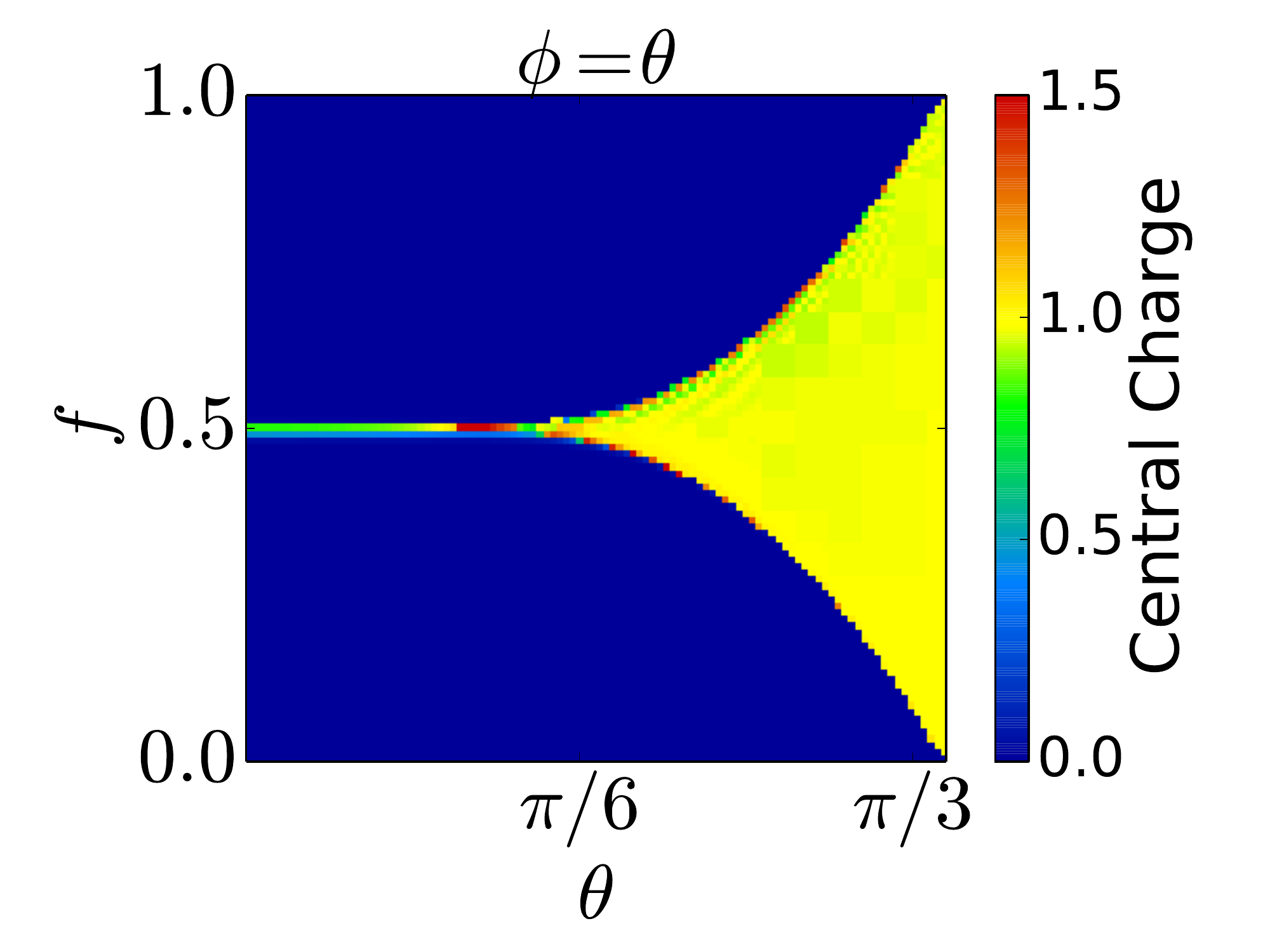} \label{phase_7}}
\caption{Three cross-sections  corresponding to 
(a) $\phi=0$ (b) $\phi=\pi/3$ and (c) $\phi=\theta$ 
of the three dimensional phase diagram, and all for $L=100.$ Topological, 
trivial, and incommensurate (IC) phases are 
identified by the central-cut entanglement entropy (color coded).
For (a) and (b) a 2D grid in 
increments of 0.01 was used to resolve fine features of the transitions.
(c) was mapped out on a 2D grid in increments of 0.05.
Point A is the transition point of the 3-state 
 Potts model, i.e. the chiral clock model for  ($\theta=\phi=0$). 
Points B and C are Lifshitz points and are 
associated with putative tricritical behavior.
The solid lines, dashed lines, and dotted-dashed lines 
indicate direct topological-trivial $(c=4/5)$ type, Kosterlitz-Thouless type, 
and Pokrovskii-Talapov type~\cite{PT} transitions respectively. The thick circularly-dotted line represents an upper bound on the region where exact parafermionic zero modes can exist~\cite{fendley2012}.
Panels (d), (e) and (f) show the corresponding central charges for cross sections (a),(b),(c) 
respectively. The IC phase is associated with central charge $c=1$ (yellow) whereas 
the critical regions close to point A have $c=4/5$ (green). 
}\label{phase}
\end{figure*}

\prlsec{Results} 
Let us now move on to discuss the results of our numerical calculations.
First, we present the full  three-parameter phase 
diagram ($f$,$\theta$,$\phi$) over the reduced domain in Fig.~\ref{phase_3d}, where we have set $J=1-f$. 
The basic topology of the phase structure is clear. We find
three distinct phases as mentioned above. The phase corresponding to largest $f$ 
values is generically the trivial phase, and the phase corresponding to the smallest $f$ values is 
generically the topological phase. They share a common/direct phase boundary between them when $\theta$ 
and $\phi$ are small. For large $\theta$ or $\phi$, an intermediate
incommensurate phase appears between the two.   

We show the central-cut EE in Fig. \ref{phase}(a),(b),(c) for several 2D cross-sections of the 3D phase diagram. These plots help to identify the gapped phases and the topology of the phase boundaries. To more clearly identify the nature of the critical regions/boundaries we also calculate the central charge via the scaling relation.
It is interesting to see that  the observed locations of the phase boundaries for cross sections 
$\phi=0$ and $\theta=\phi$ are broadly consistent with earlier 
works~\cite{ostlund1981,howes1983}, and that the topological phase itself is stable over a large part 
of the phase diagram~\footnote{By this we mean that the system remains gapped and the topological 
ground-state degeneracy is robust. We do not mean that the edge zero-modes remain exact over the entire phase range. 
See Ref.~\onlinecite{jermyn2014} for discussion on this distinction.}.

We indicate several special points on these cross sections: 
Point A in Fig.~\ref{phase_0} and Fig.~\ref{phase_6} is the transition point of the 
three-state Potts model associated with $c=4/5$~\cite{CFT}, and 
Point B and C are putative tri-critical points.  We indicate approximate locations of the 
phase boundaries with solid, dashed, or dot-dashed lines, depending on the nature of the 
phase transition, as indicated in the figure caption. 
Finite size effects were checked around specific points along the critical lines by running system sizes 
of $L$=100 to 400 on a finer grid. The locations of these lines did not change significantly 
in comparison to the resolution of our grid, except in certain regions 
which are discussed in further detail in later sections.

From the central-cut EE we see that the trivial phase is characterized by a small EE,
while the topological phase has a 
nearly uniform EE of $ \approx \ln 3$ indicating a three-fold degeneracy 
of the ground state. The change of EE is abrupt between the two phases 
as can clearly been seen in Fig.~\ref{phase_0} and Fig.~\ref{phase_6} for $\theta\lesssim\pi/4$ 
and $\theta\lesssim\pi/6$ respectively. We also verified that this transition 
is accompanied by a divergence in the second order derivative 
of the ground state energy (not shown).

The third phase in the phase diagram is the incommensurate phase.
This is a critical phase in which the correlation functions generically behave 
as $A(r) e^{(2\pi i/3) Qr}$, where $A$ decays algebraically and $Q$ is irrational. 
The oscillatory properties of the correlation functions also
manifest themselves in oscillatory behavior seen in energy gaps, which we address later. 
Although there is not an extremely sharp distinction between the central-cut EE for 
the topological and incommensurate phases, the EE  scaling with system size is markedly different. 
The former has an EE that quickly saturates to a constant value of $\ln 3$ with sub-system size, while the latter has EE 
that diverges logarithmically with sub-system size. By fitting our data to Eq.~\eqref{eq:EE}, 
we establish that the incommensurate phase is critical and its central charge is 
$c=1$ over the entire phase. 

\begin{figure*}[t!]
\subfigure[]{\includegraphics[width=0.32\linewidth]{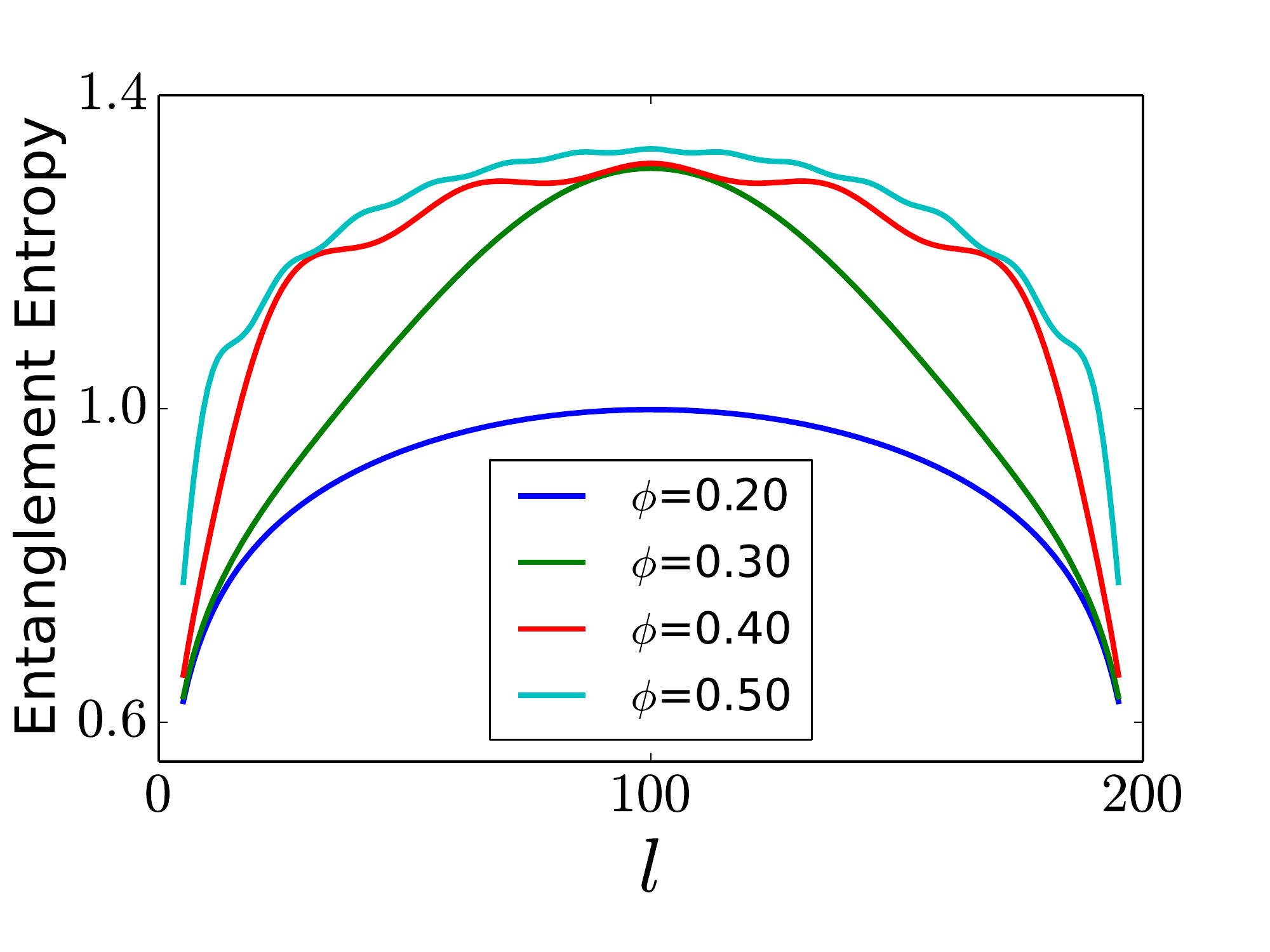} \label{entropy}}
\subfigure[]{\includegraphics[width=0.32\linewidth]{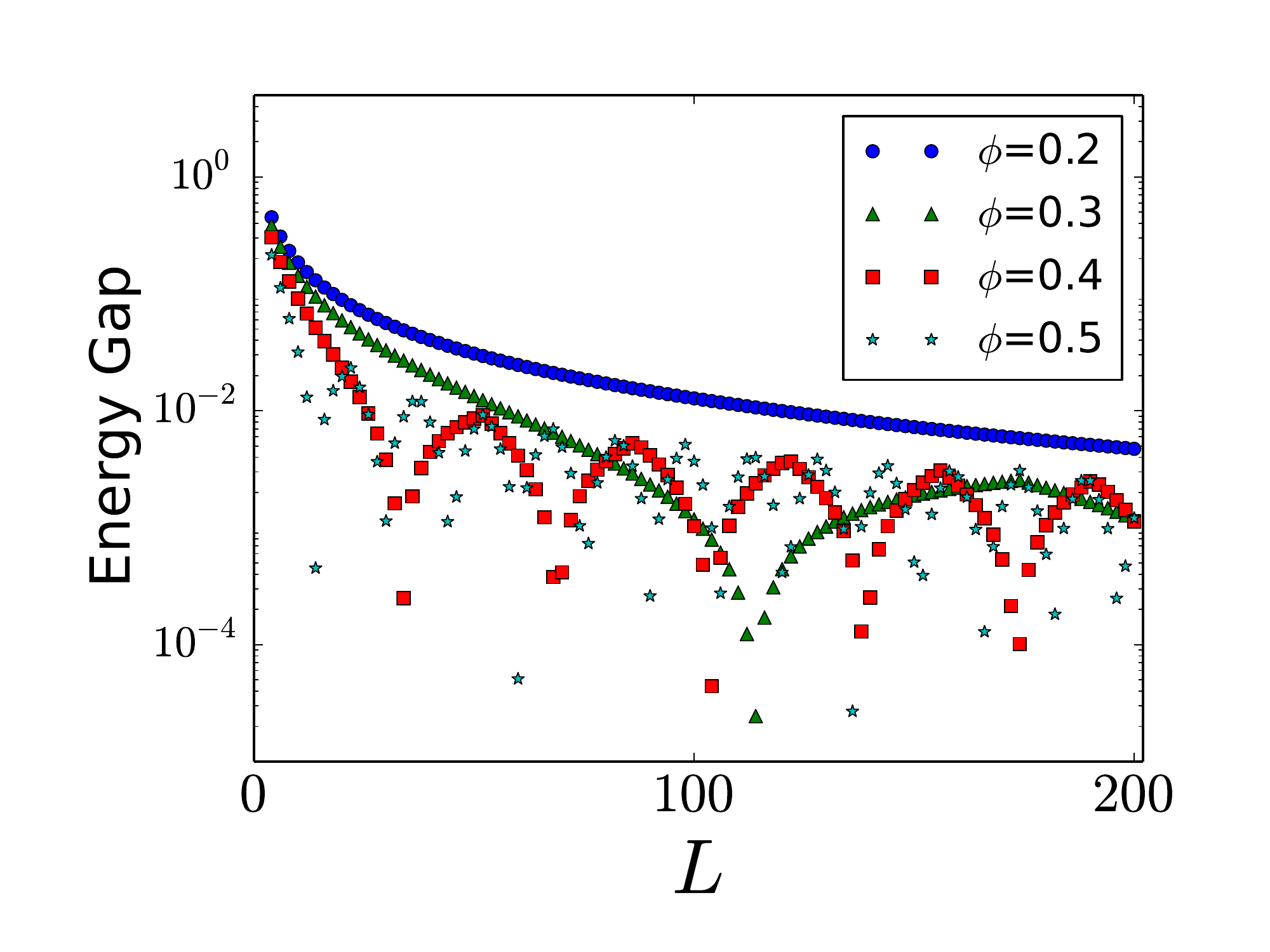} \label{gap}}
\subfigure[]{\includegraphics[width=0.32\linewidth]{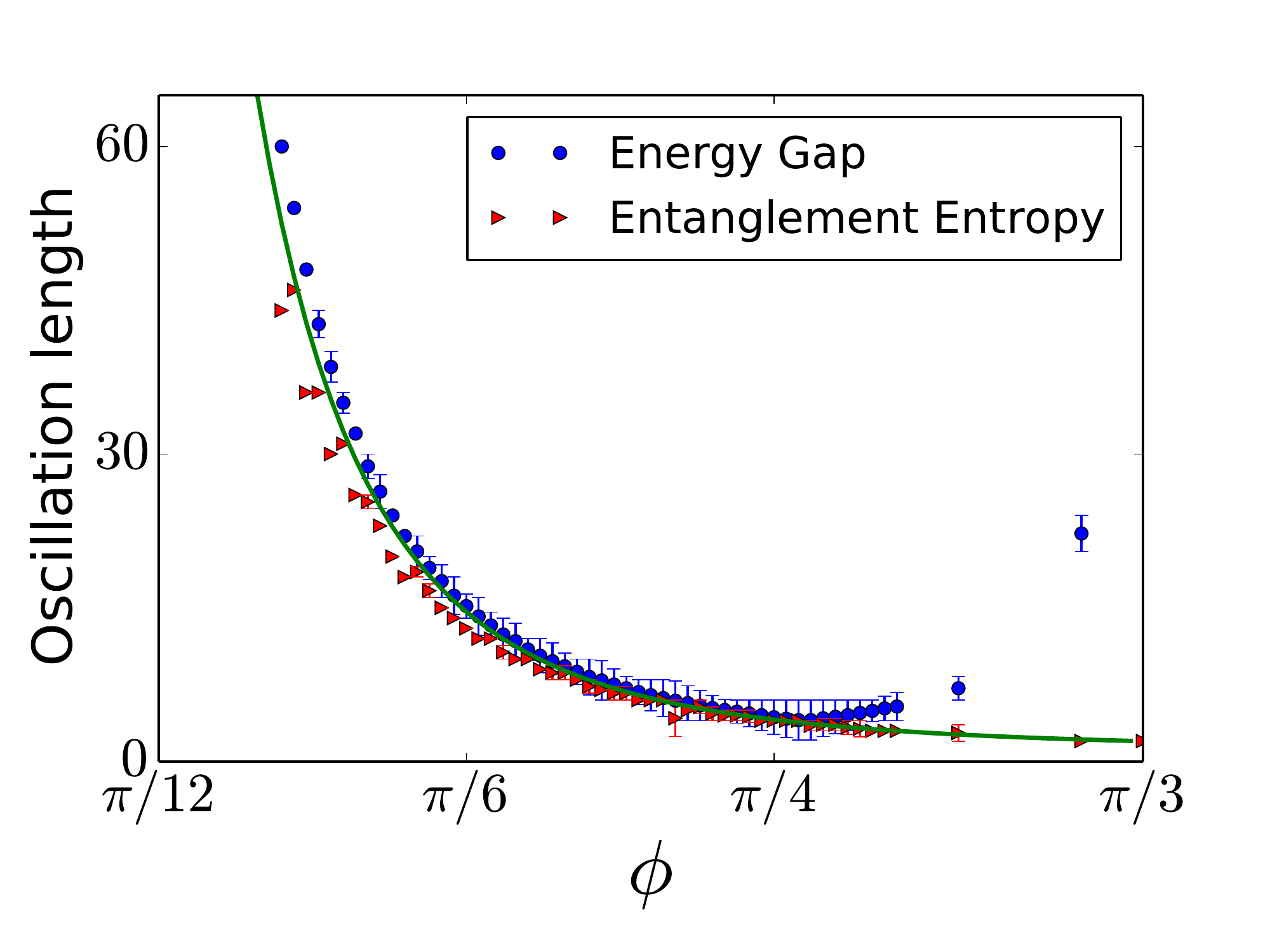} \label{oscillation}}
\caption{(Color online) Properties of the critical line at $f=J=1/2$ for various 
values of $\theta=\phi$ (a) Profile of the EE 
as a function of block size shows Lifshitz 
oscillations. We predict the oscillation length 
for $\phi<0.2$ to be larger than our system size $L=200$. 
(b) Energy gap between the ground and first excited state, 
which displays similar oscillatory behavior on varying the system size. 
(c) Characteristic oscillation lengths in the EE and energy gap, 
which are nearly identical for a large range of $\phi$. 
The (green) line is the fit with $\zeta=\phi^{-3.75}+1.16$.}\label{lifshitz}
\end{figure*}

While constructing the detailed phase diagram cross sections, 
we found that while it was easy to approximate the locations of the phase boundaries, 
we often encountered difficulties in precisely nailing down 
the central charge of the corresponding critical points.
As an example, we note the appearance 
of a few points with (apparently) high central charge, 
indicated by red color, on the direct topological-trivial phase boundary in Fig.~\ref{phase_1}. 
While in some cases there may be real physics associated to this behavior, 
we show in Appendix~\ref{app:off} that a primary source for these spurious effects is fitting to 
a region of the phase diagram that is just slightly off-criticality. We show that the 
central charge is very sensitive to the precise location of the critical point, and can easily 
give $O(1)$ errors even when only slightly tuned away from criticality, and even with reasonably large-size calculations. 

Additionally, although most phase boundaries were easily identified, 
there are three regions where difficulties arise:
(i) the trivial-incommensurate phase transition at $\phi=0$ and large $\theta$
(lower-right corner of Fig.~\ref{phase_1}),
(ii) the topological-incommensurate phase transition at $\phi=\pi/3$ and small $\theta$
(upper-left corner of Fig.~\ref{phase_3}), and (iii) the Lifshitz transition area for $f=0.5$ and $\phi=\theta\sim \pi/6$ as seen in Fig.~\ref{phase_7}.  Regions (i) and (ii)  are related by duality,
and the explanation of the numerical difficulties in these regions may have a common origin. 
To explain, we recall that the trivial-incommensurate phase transition at $\phi=0$ and large $\theta,$ i.e. region (i), is of the Kosterlitz-Thouless type~\cite{ostlund1981}.
Hence, the correlation length decays as $\exp(c(T-T_{KT})^{-1/2})$ away from the transition point~\cite{kosterlitz1973,kosterlitz1974}, and this
results in a long correlation length (compared to our system size $L=100$) for this region of the phase diagram. 
The duality indicates that  region (ii) may also be near a Kosterlitz-Thouless phase transition point.
Thus, we attribute the issues with these regions as likely artifacts due to finite size effects.  We 
elaborate further on this in Appendix~\ref{app:KT}.
The remaining region (iii) requires more discussion, to which we now turn. 


\prlsec{Lifshitz behavior} 
Let us now focus on the cross-section in Figs. \ref{phase_6}, \ref{phase_7}, which corresponds to $\phi=\theta$.
Since the system is self-dual on the line $f=J$, the trivial-topological phase boundary 
should just be the line $f=J=0.5$,  a fact verified in our numerical calculations when $\theta=\phi$ are small. 
On top of the phase diagram we also plot the function $f=[2\sin (3\phi)][1+2\sin (3\phi)]^{-1}$ (in a thick circular dotted line), which represents an upper bound on the region in which exact parafermionic zero modes are expected to exist as proven in Ref. \onlinecite{fendley2012}. The region of the phase diagram above this curve are guaranteed to not have exact parafermionic zero modes, despite still being in the topological phase with the topological ground state degeneracy.
Along the critical line $f=J=0.5$, 
$c=4/5$ at the ferromagnetic point ($\phi=\theta=0$), and $c=1$ at the antiferromagnetic point ($\phi=\theta=\pi/3$)~\cite{CFT}. 
It is \emph{a priori} unclear how the central charge transitions from $c=4/5$ to $c=1,$ i.e., is it an
abrupt jump at some transition point, or does it change incrementally in stages, or perhaps something else entirely? 
Only a few studies address this question directly: among them is the work of 
Howes et al.~\cite{howes1983} who used fermion analyses and series expansions 
to conjecture that a tricritical point connecting the ordered (topological), disordered (trivial), and incommensurate phases exists at \emph{exactly} $\phi=\theta=\pi/6$. 
McCoy et al.~\cite{Albertini1989,McCoy1990} studied the super integrable line $\phi=\theta=\pi/6$ and 
suggested a modified picture with the incommensurate phase stretched all 
the way down to the point $\phi=\theta=0$ and $f=J=0.5$. 
Our results seem to support the latter picture, as  we will further develop below. 

To address the questions posed above, we studied the critical line $f=J$ carefully. 
We observed (see Fig~\ref{entropy}) that before we reach the putative
tricritcal (Lifshitz) point at $\phi=\theta=\pi/6$, the EE starts to show oscillatory behavior~\footnote{ Note, in  
Fig.~\ref{entropy} the EE curves in the incommensurate phase are not shown 
because they overlap with the curve at $\phi=\theta=0.5.$} .
The frequency of the oscillations increases as we approach the Lifshitz point from small $\phi=\theta$,
and when further increasing $\phi=\theta$ its amplitude dies out after the system clearly enters  the incommensurate phase. Conventionally, a Lifshitz transition point of this nature corresponds to a continuously varying oscillation length, and in this case it is the length scale associated with the incommensurate order. 
Interestingly, the shapes of the EE oscillation curves match those observed recently in 1D free, and interacting, fermion systems 
near Lifshitz points where the Fermi surface is augmented by additional Fermi points~\cite{rodney2013}. Thus, our result adds to the evidence of Ref. \onlinecite{rodney2013} that these types of EE oscillations are a fingerprint of the Lifshitz-type phase transition. As an aside, we mention that the Lifshitz oscillations are only present in the EE when one uses \emph{open} boundary conditions. One can easily check this by calculating the EE for free fermions as a function of next-nearest neighbor hopping\cite{rodney2013}, but with periodic boundary conditions (shown in Appendix~\ref{app:lifshitz}).

\begin{figure*}[t!]
\subfigure[]{\includegraphics[width=0.32\linewidth]{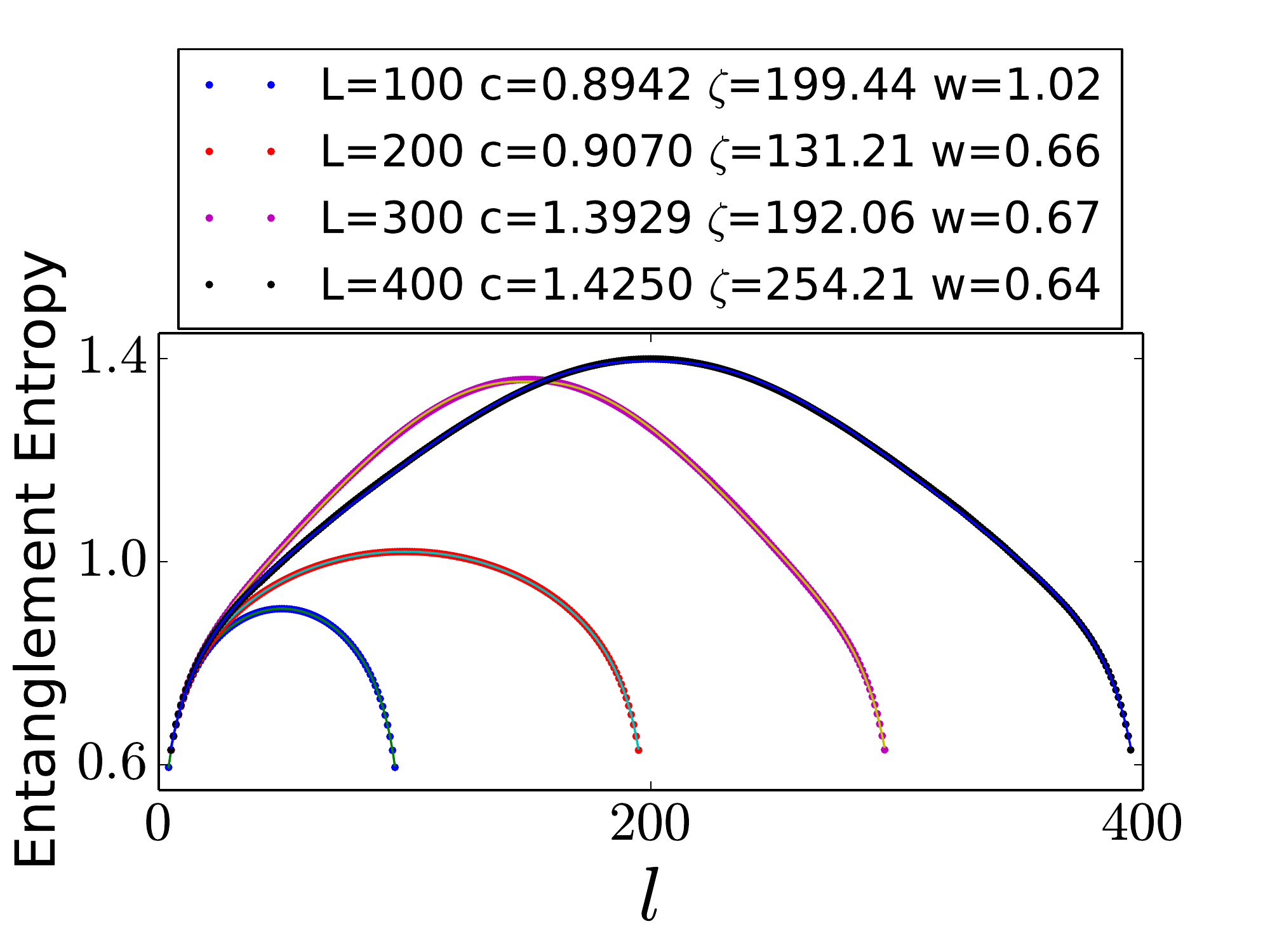} \label{oscillation_phi25}}
\subfigure[]{\includegraphics[width=0.32\linewidth]{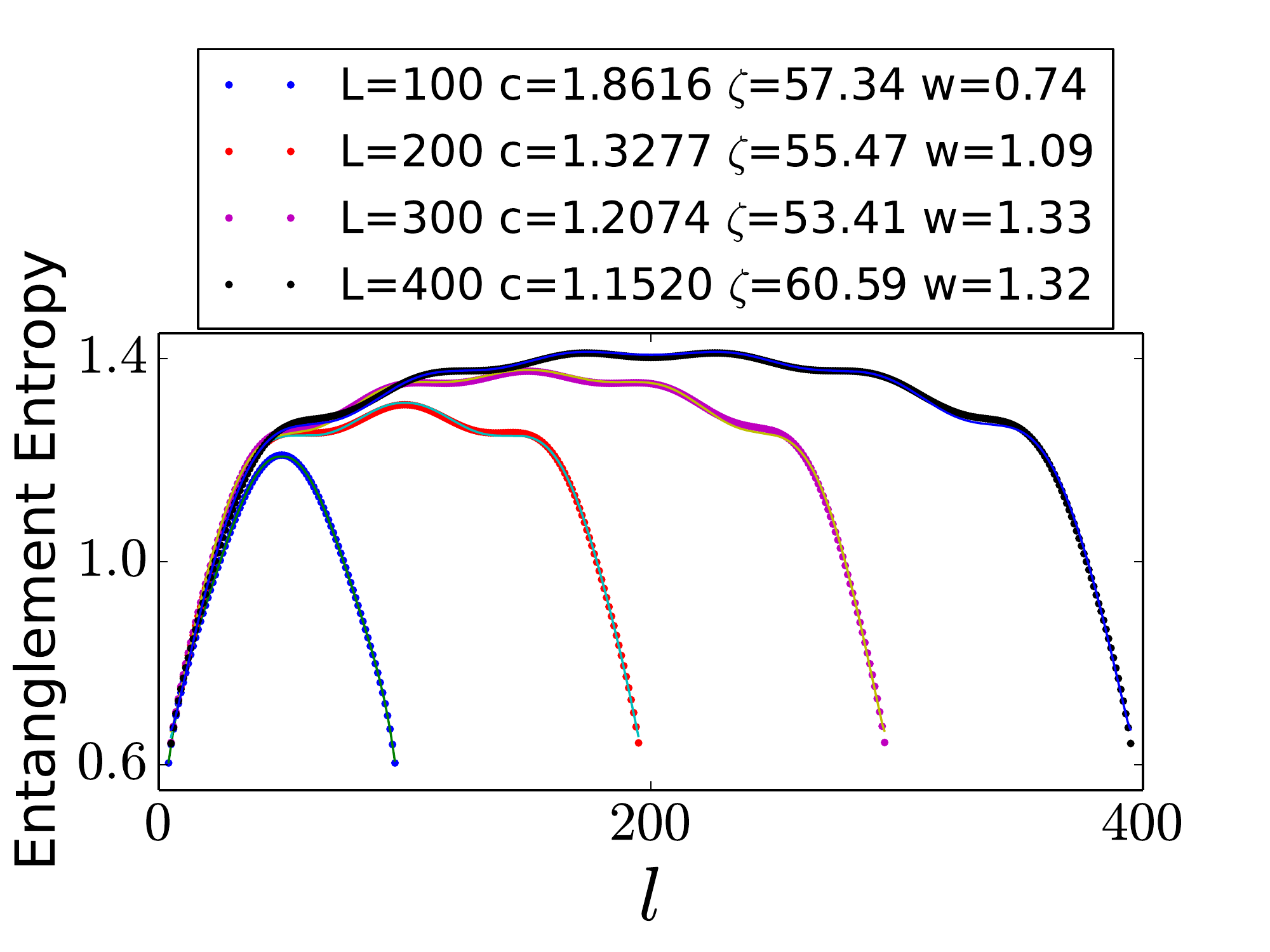} \label{oscillation_phi30}}
\subfigure[]{\includegraphics[width=0.32\linewidth]{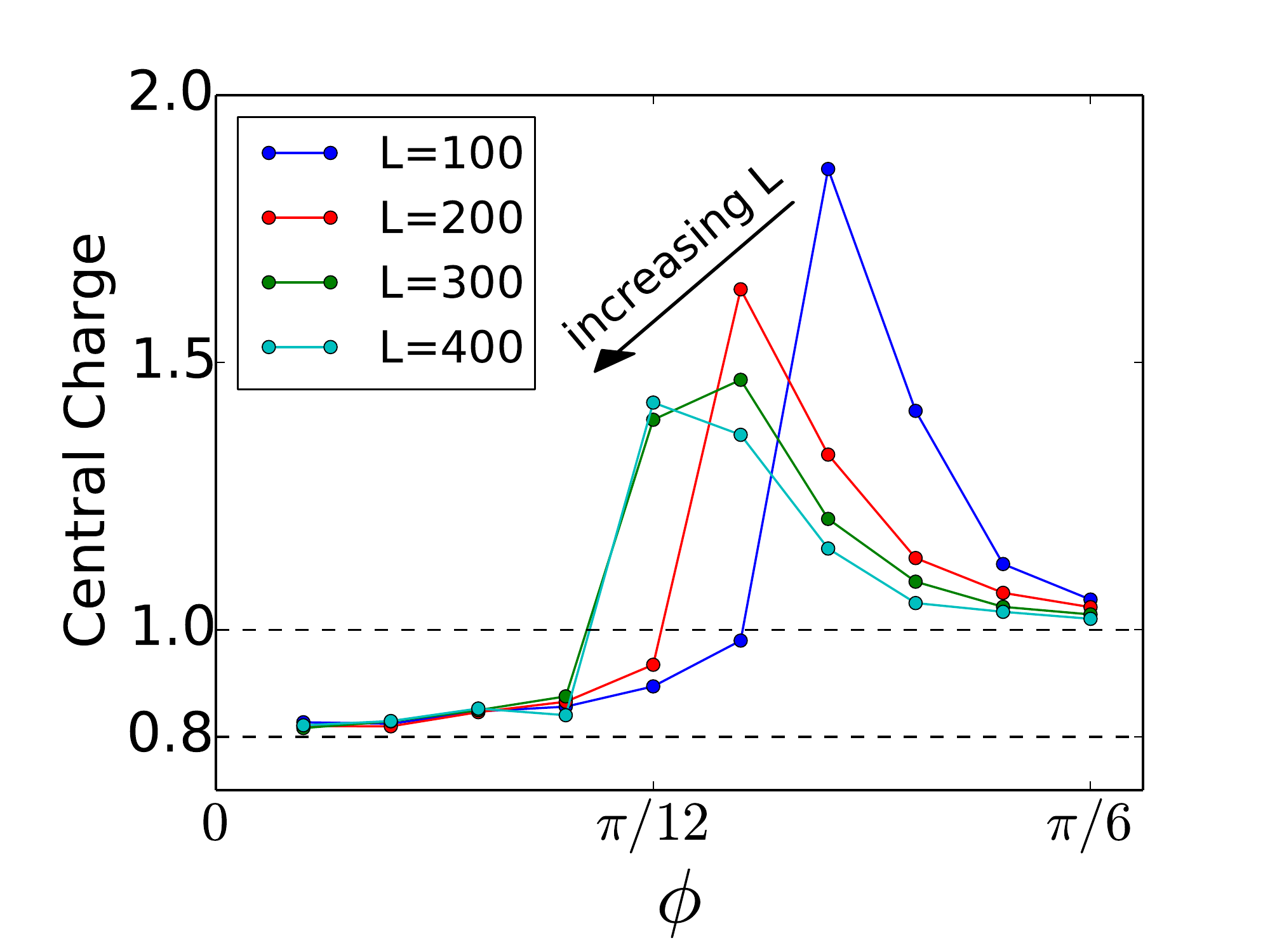} \label{oscillation_charge}}
\caption{Panels (a) and (b) show the profile of the entanglement entropy (as a function of block size) 
for various values of system size at $\phi=\theta=0.25$ and $\phi=\theta=0.35$ respectively.
Panel (c) shows the central charge obtained by fitting the entanglement entropy with the corrected
formula along the line $\phi=\theta$ and $f=J=0.5$. The two dashed lines are at $c=0.8$ and $c=1$.
The arrow indicates the trend of the peak when $L$ is increased.
}
\end{figure*}

To quantitatively study the nature of this critical regime we want to investigate the variation of the central charge. However, in the presence of oscillations in the EE, we must modify Eq.~\eqref{eq:EE} if we wish to extract the central charge. 
Empirically, the observed oscillations appear to have a similar form to those in the work Ref.~\onlinecite{Norm}, and we propose a phenomenological scaling form which can 
fit the EE with oscillations: 
\begin{equation*}
S(l)_{\textrm{cor}} = \frac{c}{6} \ln \left( \frac{L}{\pi} \sin \frac{\pi l}{L} \right) + S_0 + \frac{\cos(2\pi l/\zeta+p)}{(L/2-|L/2-l|)^w},
\end{equation*}
where the first two terms are the same as in Eq.~\eqref{eq:EE}, 
and the third term incorporates oscillations and a symmetrized damping function.
The parameter $\zeta$ is the oscillation length and $p$ is a phase factor. 
These parameters, along with the exponent 
$w,$ are free-parameters determined by fitting. 
Some representative fits are shown in Figs.~\ref{oscillation_phi25} and \ref{oscillation_phi30}, which clearly capture the sub-leading oscillations accurately.

The results of calculating the central charge from this procedure 
are shown as a function of $\phi$ 
in Fig.~\ref{oscillation_charge}. One can see that there is still an unaccounted 
for effect that leads to a peak in the central charge at a system-size dependent $\phi$ value.
More careful inspection reveals that 
the peak is located at a $\phi^*$ that corresponds to an oscillation length $\zeta\approx L/2.$
Thus, as seen in the figure, the peak location $\phi^{*}$, occurs at values closer and closer to $\phi=\theta=0$ 
when system size is increased, and all other parameters remain fixed.
Our observations indicate that the central charge converges to $c\approx 1$ 
when $\phi \ge \phi^{*}$, and $c\approx 4/5$ for  $\phi < \phi^{*}.$ This strongly suggests that the transition from $c=4/5$ to 
$c=1$ along the line $f=J=0.5$ is an abrupt one that occurs at $\phi=\theta=\phi^*$. From our numerical data it appears that $\phi^* \to 0$ as $L\to\infty.$ Hence, our data supports a scenario where there is an
immediate onset of oscillations as one tunes away from $\phi=\theta=0$ in the thermodynamic limit. 

We corroborate this by observing that oscillations are not seen in the EE 
if the oscillation length itself exceeds the system size $L.$ 
For example, for $L=200$, the oscillations are not explicitly visible for $\phi \lesssim \pi/12,$ however upon increasing the system size, with all other parameters fixed, the oscillations appear over a larger region of $\phi,$ as is shown in Fig.~\ref{oscillation_phi25}. As $\phi$ is decreased the oscillation length increases, and thus we must use larger and larger systems to observe the oscillations. 
Thus, we believe that this is evidence that, in the thermodynamic limit, the oscillations are a feature for all $\theta=\phi$ except $\theta=\phi=0$. 
An alternate scenario, which we can not rule out completely based on this numerical 
data, is that the incommensurate phase persists to 
small but non-zero values of $\theta=\phi$.
Thus, a conservative estimate of the location of the tricritical point is $0 \leq (\theta=\phi)<0.25$,
which is well below the previously conjectured location at $\theta=\phi = \pi/6$.
We aim to shed further light on this transition through larger scale simulations in future work.

Finally, we note that matching oscillations are observed in the 
splitting of the lowest two energy states (Fig.~\ref{gap}), as a function of system size. 
We can extract the characteristic length scale $\zeta$ of the oscillations from both the EE (for a given 
system length), and the energy gap (as a function of system length). Our results are shown 
in Fig.~\ref{oscillation} where a clear correlation between the two is observed for $\phi=\theta<\pi/4.$ 
The solid (green) line in Fig.~\ref{oscillation} is the fit of the oscillation 
length for $\phi=\theta<\pi/4$ to the function $\zeta=\phi^{-3.75}+1.16$. When $\phi=\theta=0$, 
the oscillation length appears to diverge, indicating that no such oscillations 
survive in the non-chiral $3$-state Potts model limit. Attempts to relax the fit with 
$\zeta=(\phi-\phi^{*})^{-\eta}+const$ (i.e. with a possibly non-zero $\phi^{*}$) 
gave $\phi^{*} \sim 0.09 $ indicating that the conjectured 
tricritical point may be in close proximity to $\phi^{*}=0$. 


\prlsec{Conclusions} 
In summary, we have mapped out the three dimensional 
phase diagram of the $Z_3$ \emph{chiral} 
clock model using the density matrix renormalization group method. 
Using the entanglement entropy (of the half-chain) 
as a diagnostic, we have been able to locate 
the phase boundaries of the various 
topological-trivial-incommensurate phase transitions.
Quantitatively, we have also been able to see the variation of the central charge along 
the various critical surfaces that divide these phases. 
Another outcome of this study is the identification of 
the Lifshitz transition using the entanglement entropy, along with an estimate of the 
location of the putative tricritical point. We discussed several 
competing qualitative scenarios for the cross section of the phase diagram in which the tricritical point has been predicted to exist. 
Our data suggests that the tricritical point (along $f=J=1/2)$ 
is not at $\phi=\theta=\pi/6$: rather we find it to be shifted
to a much smaller value in the range $0 \leq \theta=\phi <0.25 $. 

Finally, our results must be viewed in a broader 
context as providing further confirmation of the stability of 
the parafermionic topological phase to chiral interactions, 
over a wide range of parameters. We expect a further study 
of this and related models to elucidate the 
conditions under which these phases can be practically 
realized. 

\prlsec{Acknowledgement}
We thank P. Fendley and G. Ortiz for discussions. HJC was supported by SciDac 
grant DOE FG02-12ER46875. NMT was supported by DOE DE-NA0001789.  Computer time was provided by XSEDE, supported by the National Science Foundation Grant No. OCI-1053575, the Oak Ridge Leadership Computing Facility at the Oak Ridge National Laboratory, which is supported by the Office of Science of the U.S. Department of Energy under Contract No. DE-AC05-00OR22725 and Taub campus cluster at UIUC/NCSA. TLH is supported by the US National Science Foundation under grant DMR 1351895-CAR.


\appendix
\section{Properties of the $3$-state chiral clock model}\label{app:duality}

The Hamiltonian in Eq.~\eqref{eq:H3} has the same properties when 
 either of the two phases $\theta, \phi$ are shifted by multiples of $\frac23\pi$. To show this we can see that  
 the transformation 
\begin{equation}
    \theta' \rightarrow \theta + \frac{2n\pi}{3} \;\;\;\; \phi' \rightarrow \phi + \frac{2m\pi}{3} \;\;\;\;
\end{equation}
changes the Hamiltonian to: 
\begin{equation}
	H_3 = -f \omega^{-n} \sum_{j=1}^L \tau_j^{\dagger} e^{-i\phi} - J \omega^{-m} \sum_{j=1}^{L-1} \sigma_j^{\dagger} \sigma_{j+1} e^{-i\theta} + \textrm{h.c.}
\end{equation}
Then we can redefine the operators:
\begin{equation}
	\tau' = \omega^{-n} \tau \;\;\;\; \sigma'_{2j}=\omega^{-m} \sigma_{2j} \;\;\;\; \sigma'_{2j+1}=\sigma_{2j+1}.
\end{equation}
This new set of operators preserves the properties $\tau^3=\sigma^3=I$, 
$\sigma\tau=\omega\;\tau\sigma$, where $\omega=e^{2\pi i/3}$. After this redefinition we end up with a Hamiltonian with the same form as the original.

Additionally, the transformation that flips the signs of the two phases at the same time, i.e.,
\begin{equation}
    \theta' \rightarrow - \theta \;\;\;\; \phi' \rightarrow - \phi\;\;\;\;
\end{equation}
changes the Hamiltonian to,
\begin{equation}
        H_3 = -f \sum_{j=1}^L \tau_j e^{-i\phi} - J \sum_{j=1}^{L-1} \sigma_j \sigma_{j+1}^{\dagger} e^{-i\theta} + \textrm{h.c.}
\end{equation}
Here we can redefine the operators as
\begin{equation}
        \tau' = \tau^{\dagger} \;\;\;\; \sigma'=\sigma^{\dagger}
\label{eq:dagger_transform}
\end{equation}\noindent to recover the form of the original Hamiltonian.

We can also just  flip the sign of just one of the phases, say 
$\phi' \rightarrow - \phi$, and then the redefinition:
\begin{equation}
        \tau'_{j} = \tau_{-j} \;\;\;\; \sigma'_{j}= \sigma_{-j}
\label{eq:j_minusj}
\end{equation}
leaves the Hamiltonian unchanged. If instead we flipped the sign of 
$\theta,$ will need a transformation that involves 
both Eqs.~\eqref{eq:dagger_transform} and ~\eqref{eq:j_minusj}.

Finally, we can consider the duality transformation:
\begin{equation}
\mu_{j+\frac12} = \prod_{k=1}^j \tau_k, \quad \nu_{j+\frac12} = \sigma_{j}^{\dagger} \sigma_{j+1}.
\end{equation}
These dual operators satisfy  $\mu^3=1$, $\nu^3=1,$ and $\mu\nu=w\;\nu\mu$. The dual Hamiltonian is then
\begin{equation}
H_3^{\textrm{dual}} = -J \sum_{j=1}^{L-1} \nu_{j+\frac12}^{\dagger} e^{-i\theta} - f \sum_{j=1}^{L} \mu_{j-\frac12}^{\dagger} \mu_{j+\frac12} e^{-i\phi} + \textrm{h.c.}
\end{equation}
Comparing with the original Hamiltonian, the dual Hamiltonian returns to the original 
form if we exchange $\theta$ and $\phi$, and at the same time $J$ and $f$.


\section{Extracting Central Charge Near Critical Points}\label{app:off}
%
When performing the fit to EE data obtained from a finite size system, and for a
point in parameter space that is close to (but not at)
a critical point, it is often difficult to obtain a reasonable estimate of the central charge. 
One possible explanation is that,
when the system size is smaller than the correlation length, 
the fit to Eq.~\eqref{eq:EE} may appear to be good, 
but the central charge obtained from the fit may 
not  match the actual central charge of the nearby critical point. 
This is not unique to our model, and we were also able to observe this effect for free Dirac fermions with a tunable mass term as shown below.
Eventually, if the system is tuned off criticality, and when the system size is larger than the correlation length, 
the EE will saturate and hence reveal the gapped phase. 

\begin{figure}[htpb]
\centering
\includegraphics[width=\linewidth]{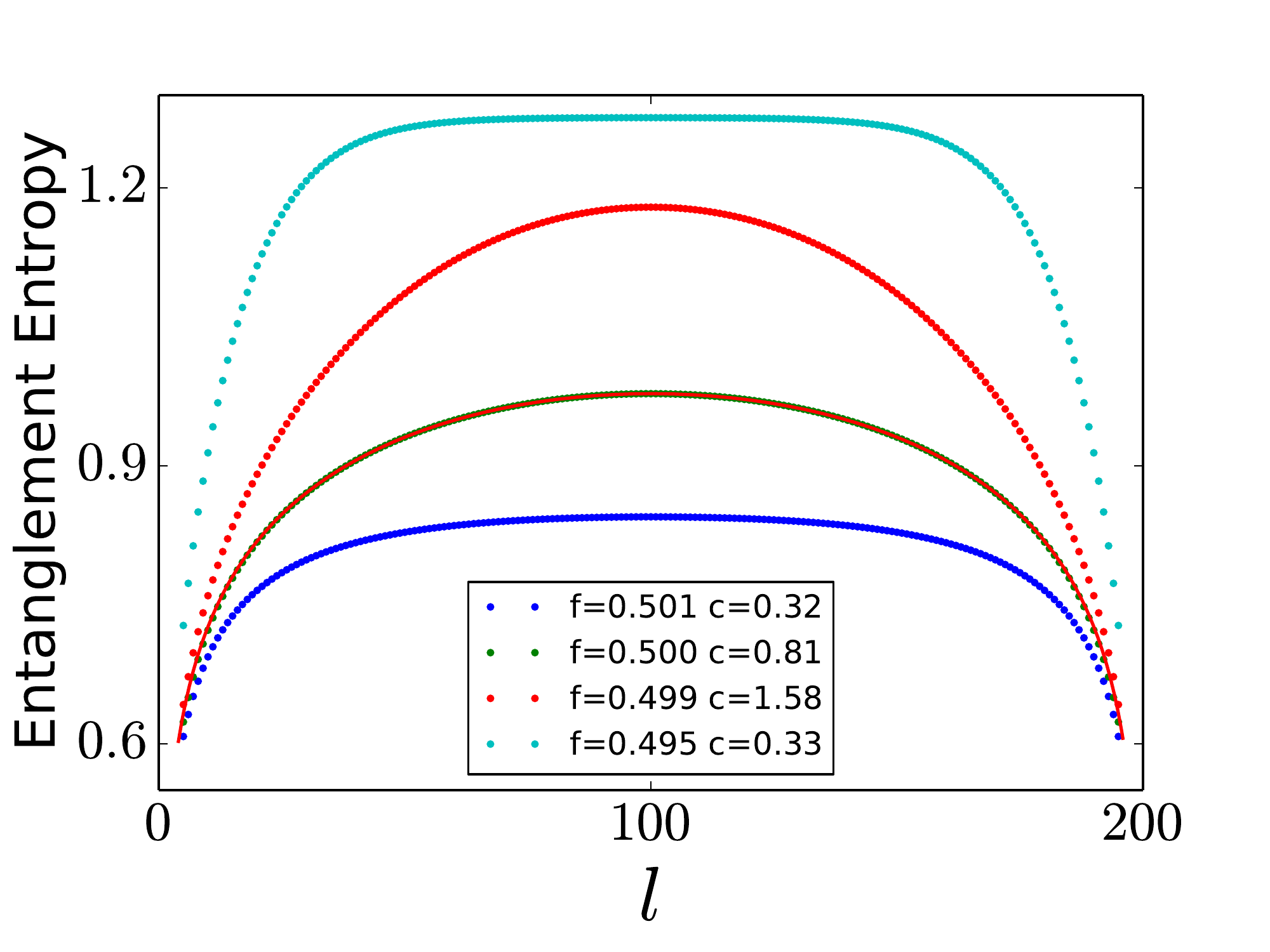}
\caption{The EE as a function of the subsystem size $l$ at 
$\phi=\theta=0$ and several different $f$ close to or at the critical point ($f=0.50$). 
From the highest curve to the lowest one, the corresponding $f$ is
0.495, 0.499, 0.500, and 0.501.
The central charges obtained from the fitting are shown in the legend. 
For $f=0.495$ and $f=0.501$, a plateau in the EE is seen indicating a gapped phase. 
For $f=0.499$, an apparent critical phase is seen which is attributed to an artifact 
of finite size effects.
 } \label{fit}
\end{figure}

To provide an example of such behavior, we refer to known analytic results 
that the central charge should be $4/5$ at ($f=J=0.5$, $\phi=\theta=0$), and zero for all 
other $f$ at $\phi=\theta=0$. 
In Fig.~\ref{fit}, we show that at the critical point $f=J=0.5$, the central charge is $c=0.81\pm0.01$, 
close to the analytical result. However, when we are slightly away from this point,
say $f=0.499$, the system still appears critical with an (apparent) 
central charge of $c=1.58$, much larger than the expected value of $0.80$.
On going slightly further away, $f=0.495$, a plateau in the EE profile is seen consistent with our expectation 
of a gapped phase. Thus, the fitting procedure produces misleading results in the neighborhood of the critical point, and can make it difficult to determine the central charge for critical points in which the position of the point is not known to extremely high accuracy.

\subsection{Near Critical behavior}\label{app:FF}
To further confirm our discussion above we performed similar calculations for 1D gapless Dirac fermions using exact diagonalization.
We use the 2-band free-fermion lattice Dirac model as the test model
\begin{displaymath}
\begin{array}{rl}
H =&- \sum_n \left( i c^{\dagger}_{n+1,\uparrow} c_{n,\downarrow} + i c^{\dagger}_{n+1,\downarrow} c_{n,\uparrow} +h.c.\right)\\
&- \sum_n \left( c^{\dagger}_{n+1,\uparrow} c_{n,\uparrow} - c^{\dagger}_{n+1,\downarrow} c_{n,\downarrow} +h.c.\right) \\
&+ (2-m) \sum_n \left( c^{\dagger}_{n,\uparrow} c_{n,\uparrow} - c^{\dagger}_{n,\downarrow} c_{n,\downarrow} \right)
\end{array}
\end{displaymath}
This model is gapless at $k=0$ if m is zero, and the critical point should have a central charge of 1. If m is tuned away from zero the system exhibits an energy gap of the size 2m. For our entanglement calculations the system was filled to half filling, such that when it is gapless, the filling hits exactly at the Dirac node, and if it is gapped, the filling includes all the states in the lower band. In this model the correlation length is controlled  by the scale 1/m (with units it would be $\hbar v_F/m$ but $\hbar$ and $v_F$ are effectively unity for our model).

To compare closely with our DMRG results we fit the central charge of this model using entanglement scaling with open boundary conditions. When gapless, we find the central charge to 2 or 3 digits of accuracy. For example, we find c=1.006 when the chain is of length 400.  In addition to calculating the scaling law over the entire chain we can improve the fit by taking symmetric cuts around the center of the chain which reduces the edge effects.    We get slightly improved accuracy for ranges such as 120-280, i.e., c=1.004. If we increase system size to L=500 and fit over 120-380 we find c=1.003. 

\begin{table}[ht]
\centering
\begin{tabular}{c | c c c}
 & 40-360 & 120-280 & 40-80 \\ \hline
$L$=300 & 1.0198 & 1.0202 & 1.0199 \\
$L$=400 & 1.022 & 1.024  & 1.021 \\
$L$=500 & 1.0243 & 1.0273 & 1.0219. 
\end{tabular}
\caption{The central charge obtained by fitting from different region of the system and different system size $L$. The mass gap is set to be $m$=1/10000}
\label{table4}
\end{table}

Now let us perturb the system slightly away from the critical point. For this test we turn on a gap size of m=1/10000 as a start.  As an estimate, this should give a correlation length of $\xi$ =10000 sites. For system size 400, if we fit from 40-360, we find c=1.022; if we fit from 120-280 we find c=1.024. If we try to fit a different range, e.g., 40-80 we find c=1.021. Either way, the result is already 1\% different than the gapless case even for this tiny gap (compared with the bandwidth). Next we repeated the same 3 fits for L=300 and we find c= 1.0198, 1.0202, 1.0199. And then for L=500 and find c=1.0243,1.0273,1.0219. These results are summarized in Table.~\ref{table4}. We observe that the fits get worse when we increase the system size, and when we fit over the region restricted mostly to lie over the center. The latter result may be expected since the scaling function varies most slowly over the center. The fact that the fits get worse as we increase system size is most likely just an indicator that there is a finite correlation length and that the critical scaling form will eventually break down. For additional tests we also fit the central charge for larger (but still very small) mass gaps with $m=1/1000$ and $m=1/100$ in Table.~\ref{table3} and Table.~\ref{table2} respectively.

\begin{table}[ht]
\centering
\begin{tabular}{c | c c c}
 & 40-360 & 120-280 & 40-80 \\ \hline
$L$=300 & 1.1374 & 1.1633 & 1.1230 \\
$L$=400 & 1.1699 & 1.2082  & 1.1372 \\
$L$=500 & 1.2014 & 1.2499 & 1.1473
\end{tabular}
\caption{The central charge obtained by fitting from different region of the system and different system size $L$. The mass gap is set to be $m$=1/1000}
\label{table3}
\end{table}

\begin{table}[ht]
\centering
\begin{tabular}{c | c c c}
 & 40-360 & 120-280 & 40-80 \\ \hline
$L$=300 & 1.8811 & 1.7267 & 1.9246 \\
$L$=400 & 1.7563 & 1.4033 & 1.9438 \\
$L$=500 &  1.5701 & 1.0878  & 1.9385 \\
$L$=600 & 1.3874 &  0.84507 & 1.9273
\end{tabular}
\caption{The central charge obtained by fitting from different region of the system and different system size $L$. The mass gap is set to be $m$=1/100}
\label{table2}
\end{table}

We see that when we are tuned near, but not at, the critical point the best fits in the gapped case seem to come from smaller system sizes, and over ranges which do not include the flat middle portion of the scaling range nor the far tails of the scaling range. The unfortunate thing is that once we are a bit further away from the critical point this optimized fitting pattern no longer works. In this case none of the fitting regimes we used give accurate results because the system begins to reveal its gapped nature. We do find something close to $c=1$ when $m=1/100$ and $L=500$  (Table.~\ref{table2}), but this seems accidental since we tested it for L=600 and got a worse results. From this data we would claim that for the Dirac model when the central charge differs by 20\% from its expected value then we are too far away from the critical point to do any fitting and should claim that it is not critical. In fact for a system size of 500 and mass gap of $m=1/1000$ the fitted values are closer to $6/5$ instead of $1$ and could easily lead to misidentification of critical points in models where their location is not known exactly. 

\begin{figure}[htpb]
\centering
\includegraphics[width=\linewidth]{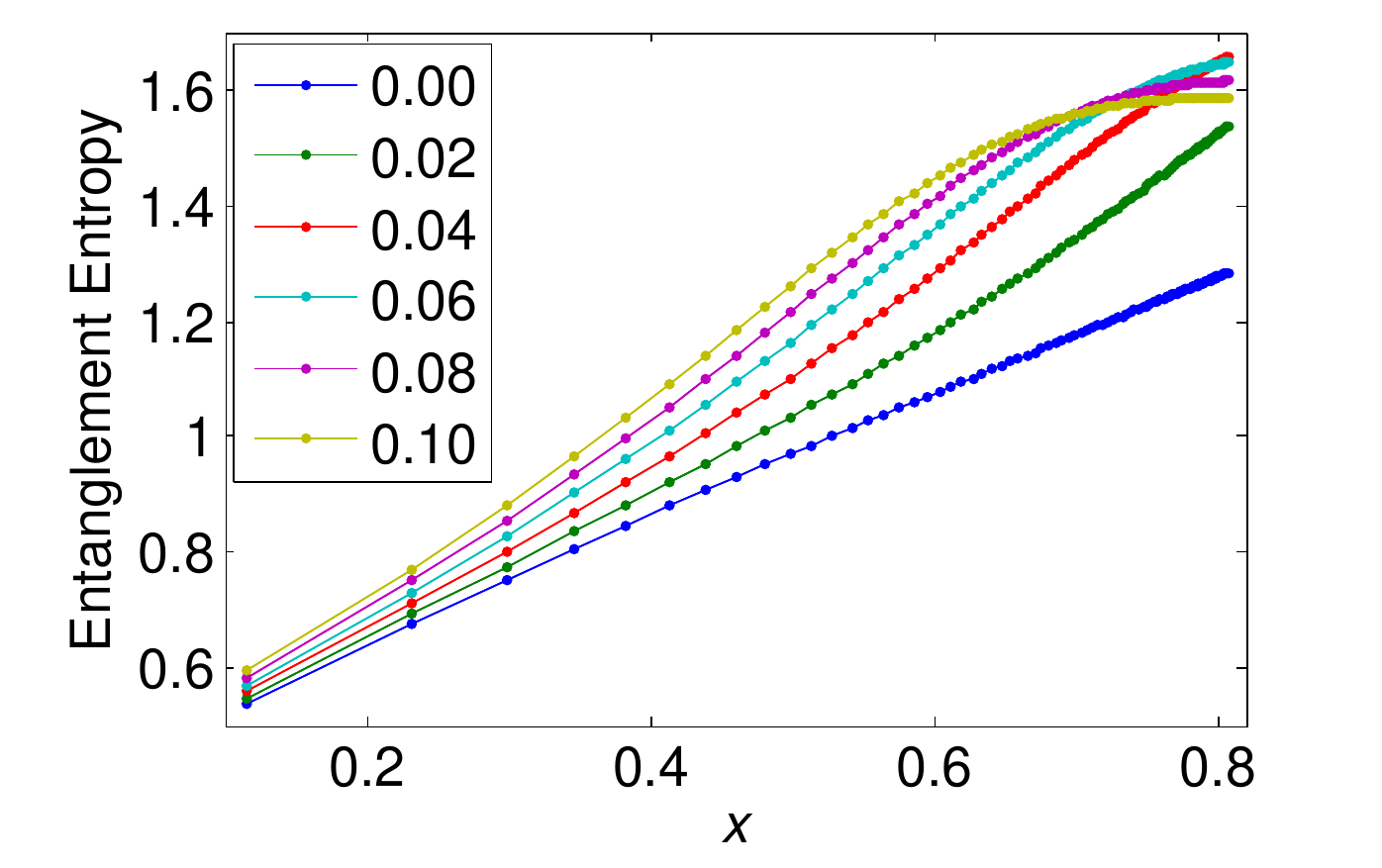}
\caption{The transformed curve of entanglement entropy. Here $x$ is defined as $\ln(\frac{L}{\pi}\sin\frac{\pi l}{L})$, where $l$ is the block size and the system size $L=200$.  For small $x$ the curves are ordered by increasing the mass gap (lowest mass is the lowest curve). } 
\label{entvsmass}
\end{figure} 

\begin{figure}[htpb]
\centering
\includegraphics[width=\linewidth]{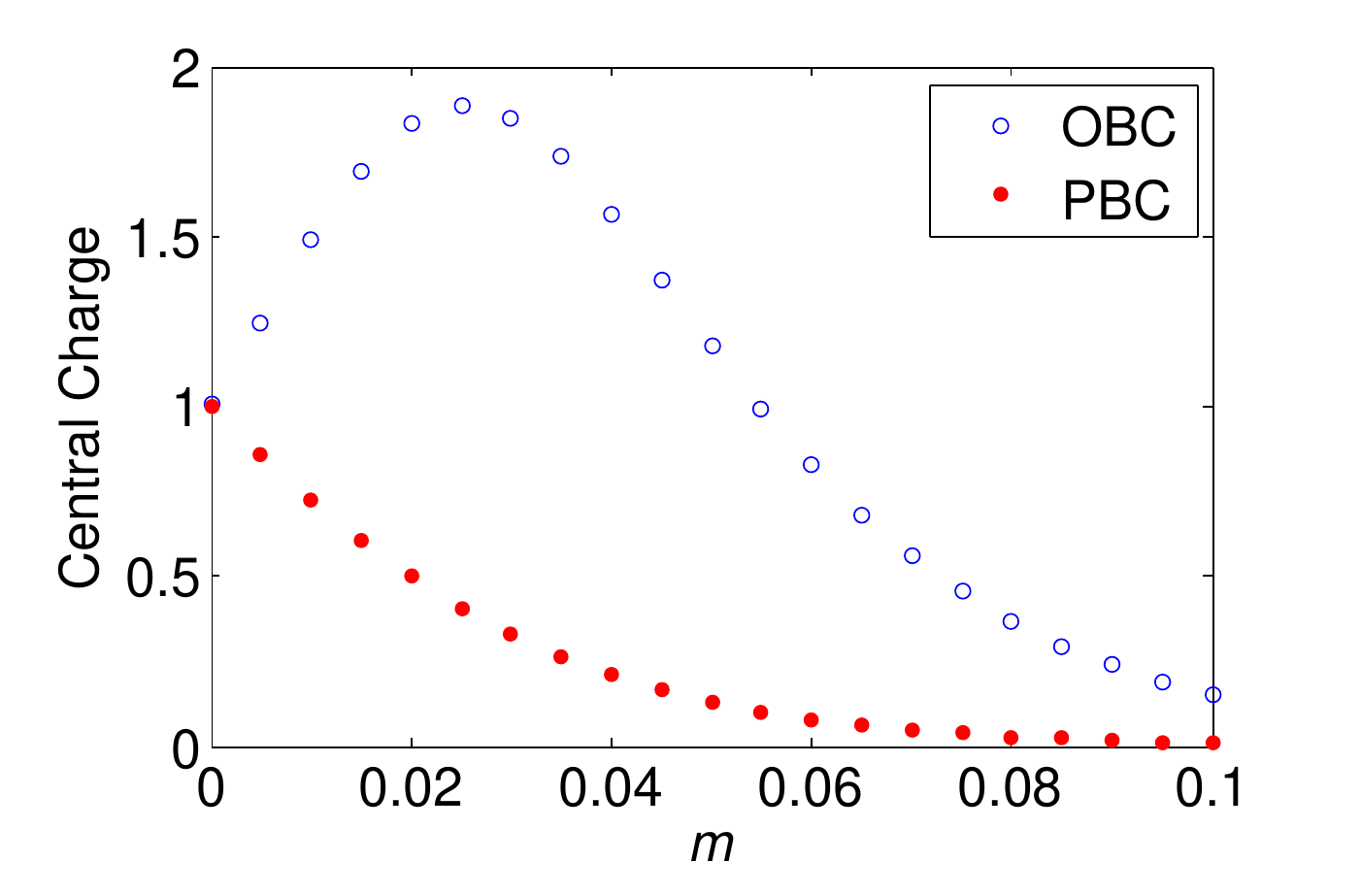}
\caption{The central charge obtained by fitting the entanglement entropy from site 40-160 for a 200 sites chain. The blue open circles are for open boundary conditions and the red dots are for periodic boundary conditions.} 
\label{chavsmass}
\end{figure}

As a possible diagnostic we plot the entanglement entropy as a function of $x=\ln(\frac{L}{\pi}\sin\frac{\pi l}{L})$, where $l$ is the sub-block size. The slope of the entanglement entropy vs $x$ should be interpreted as $c/6$. The only feature that could be used as a diagnostic is that if the transformed curve has a decrease in slope then we are definitely too far away from a critical point to fit properly as can be seen in Fig.~\ref{entvsmass}. The final two curves have clear decreases in slope as we move far away from criticality. Note that all these artificially high central charges only occur when we use open boundary conditions. As can be seen in Fig.~\ref{chavsmass}, the central charge first goes up then drops for open boundary conditions when we tune the system away from criticality. However, it decreases monotonically for periodic boundary conditions.

\section{Kosterlitz Thouless transition in the $3$-state chiral clock model phase diagram}\label{app:KT}
In the main text, we studied several 2D cross sections of the 3D $(f,J=1-f,\theta,\phi)$ 
phase diagram of the chiral clock model. The 2D cross sections corresponding to 
$\phi=0$ (see Figs.~\ref{phase_0} and \ref{phase_1}) and $\phi=\pi/3$ 
(see Figs.~\ref{phase_2}  and \ref{phase_3}) showed some regions 
whose phase boundaries could not be located.
This was attributed to finite size errors, which we now address.

\begin{figure}[htb]
\centering
\subfigure[]{\includegraphics[width=\linewidth]{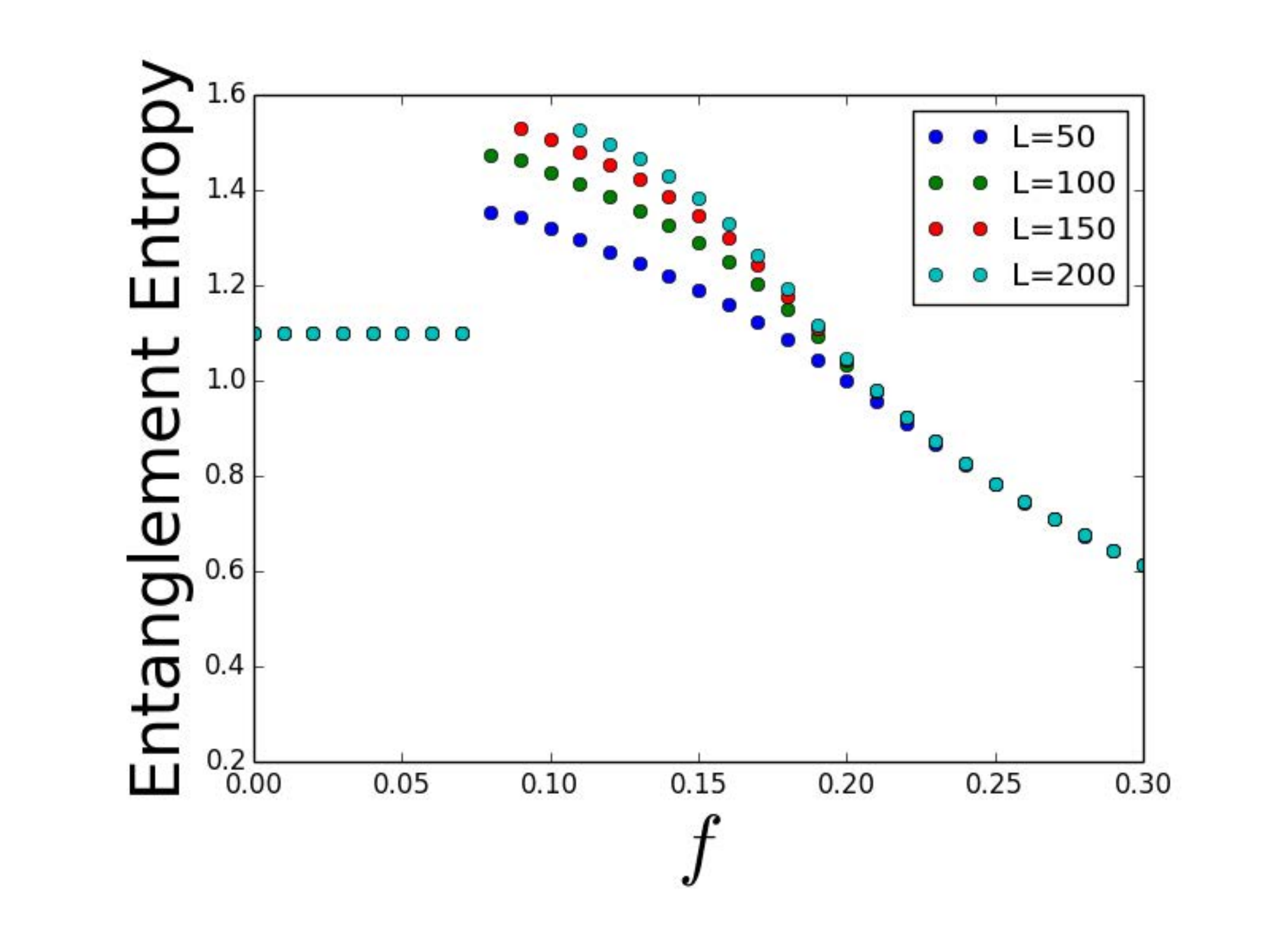} \label{entropy_01}}
\subfigure[]{\includegraphics[width=\linewidth]{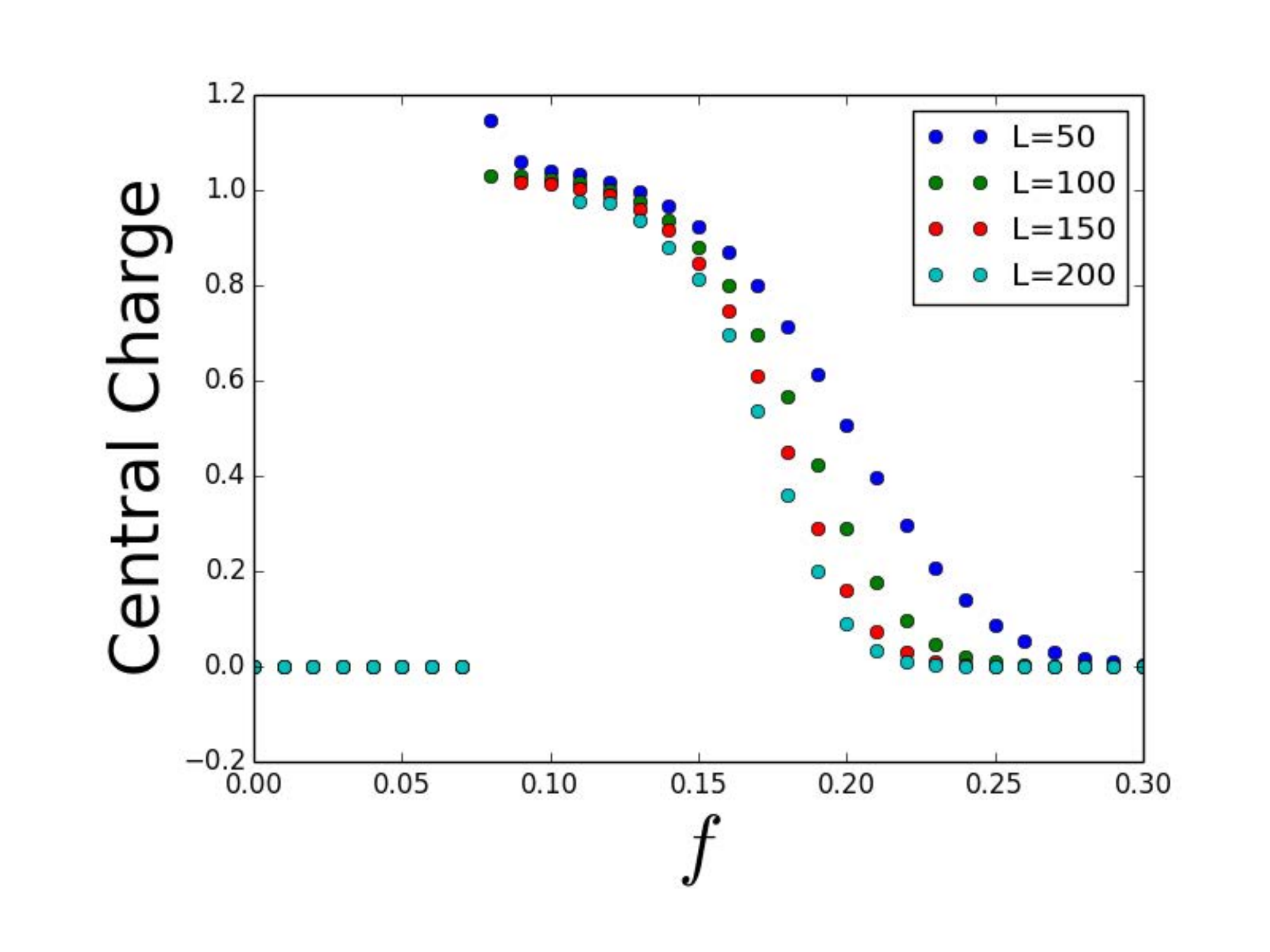} \label{charge_01}}
\caption{Panel (a) shows the entanglement entropy 
(for the central cut), as a function of $f$ for $\theta=1.00$.
The EE increases for larger sizes of the system for $f$ from 0.07 to 0.17,  
indicating a critical phase at this region.
(b) shows the corresponding central charge calculated for various system sizes.
The change of the central charge becomes sharper for larger systems.}
\label{phi0}
\end{figure}

\begin{figure}[htpb]
\centering
\subfigure[]{\includegraphics[width=\linewidth]{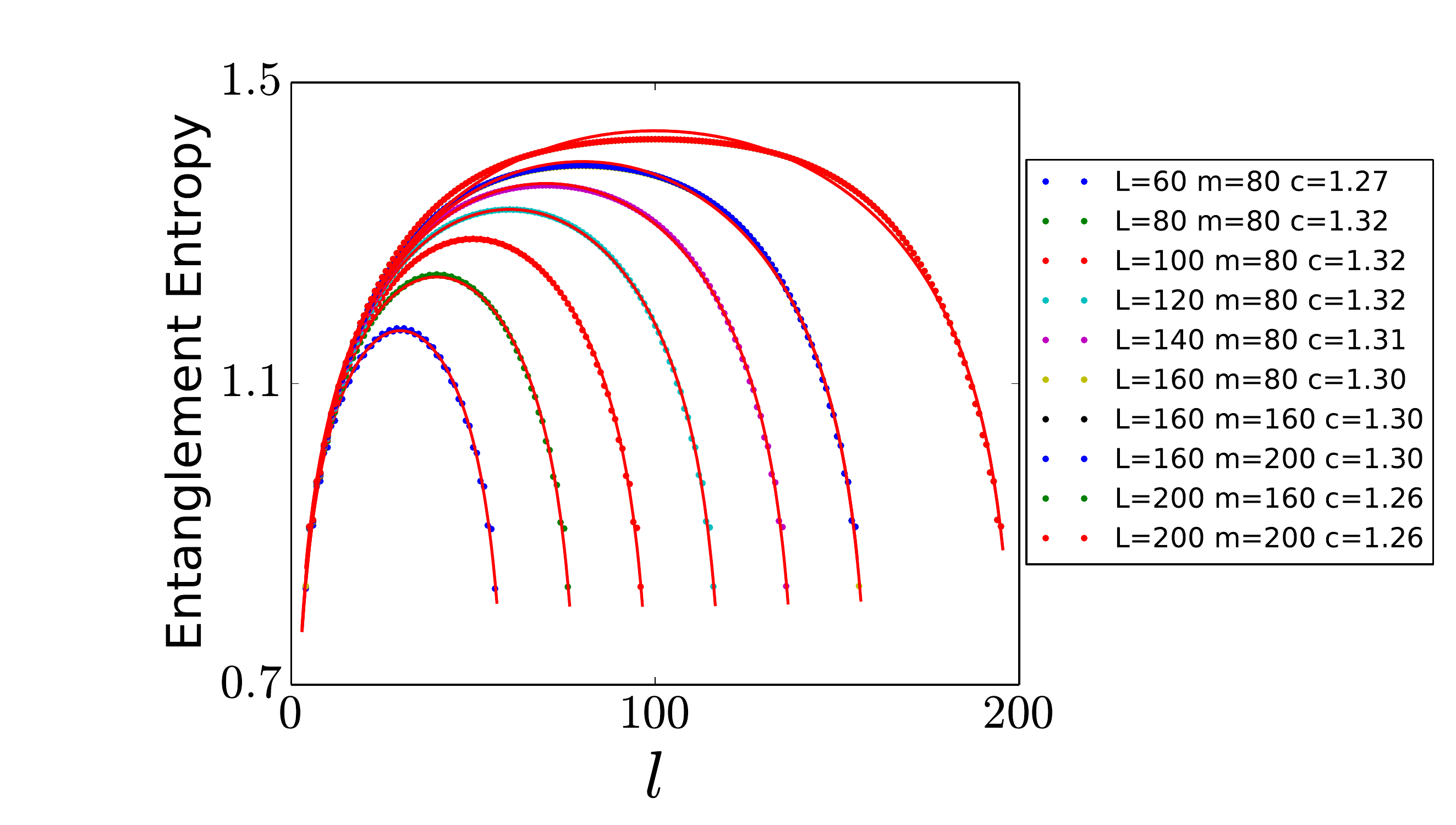} \label{entropy_bump1}}
\subfigure[]{\includegraphics[width=\linewidth]{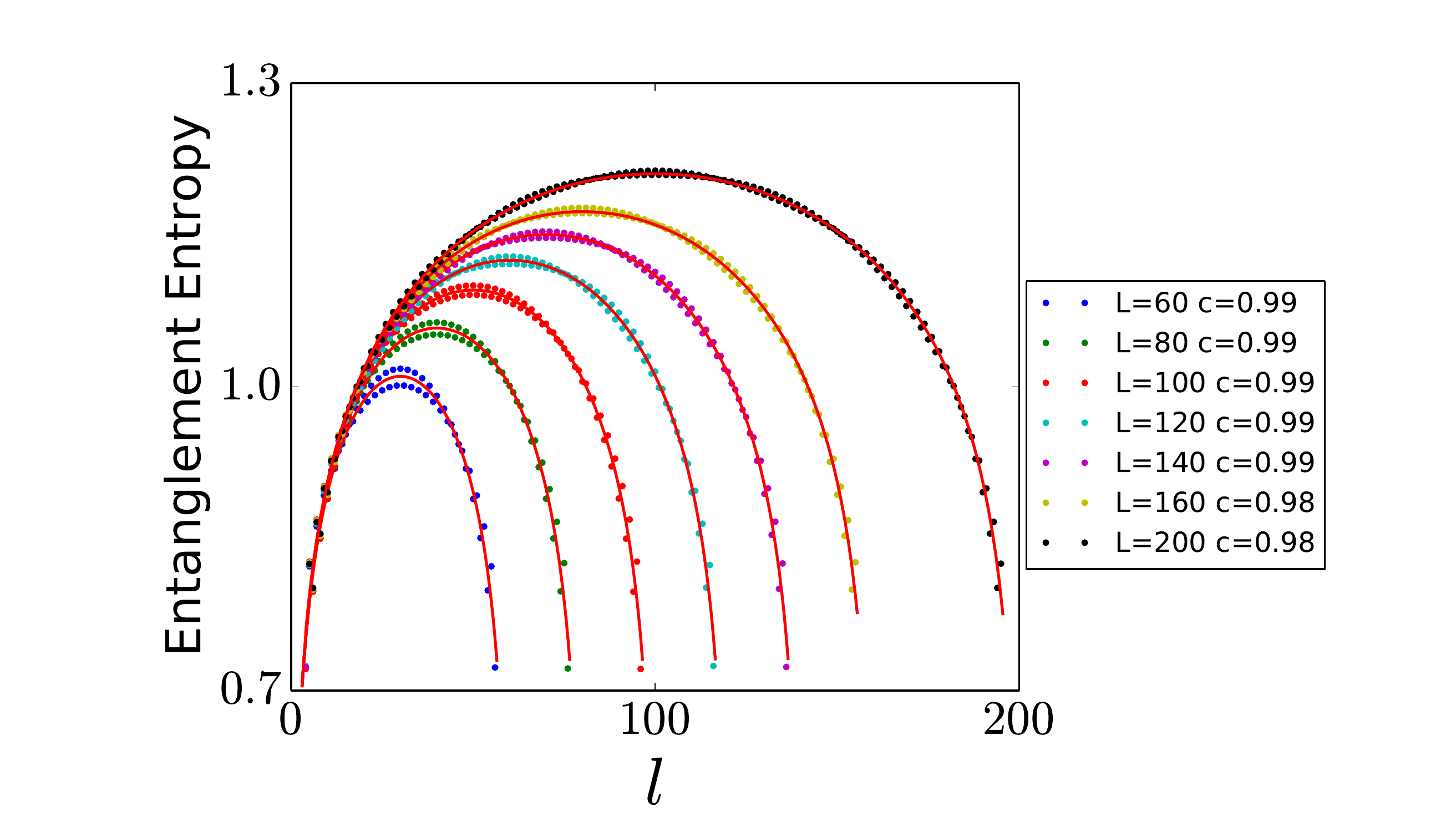} \label{entropy_bump2}}
\caption{The profile of the entanglement entropy as s function of block size $l$ 
at $\phi=\pi/3$, $\theta=0$ and (a) $f=0.8$ (b) $f=0.9$ for different system size. 
The continuous lines are the fit to the DMRG data.} 
\label{entropy_bump}
\end{figure}

We first discuss the features seen in 
Figs.~\ref{phase_0} and \ref{phase_1},
 i.e., the cross section for $\phi=0$.
For small $f$ and large $\theta,$ the phase transition 
between the topological and trivial phase is indirect: it is 
mediated by the incommensurate phase. To establish the fact that the incommensurate region 
is of non-zero extent, we performed finite size analyses on both the entanglement entropy and central charge 
as is shown in Figs~\ref{entropy_01} and \ref{charge_01} This extent is found to be 
from $f \approx 0.07 $ to $ f \approx 0.15 $.  
We find that the central charge of the trivial-incommensurate transition is consistent with that of the Kosterlitz-Thouless (KT) type, i.e., $c=1$ ~\cite{ostlund1981}. 

Because of the duality in the Hamiltonian (Eq.~\ref{eq:H3}), 
the phase diagram is symmetric with respect to the line 
$f=J=0.5$, $\phi=\theta$. Thus, the above mentioned 
phase transition is dual to the incommensurate-topological 
phase transition, for large $\phi$ and small $\theta$. 
That is to say, the region with the smooth change of the central charge in 
the lower-right corner of Fig. \ref{phase_1} is dual to the (red) region 
in the upper-left corner of Fig. \ref{phase_3}. This region, 
being near the KT phase transition point is also plagued 
by finite size errors: the correlation length is long compared with 
the system size ($L=100$). 

To test this assertion, we studied the (apparently) large central charge that was calculated near the 
critical region, as is shown in Fig.~\ref{entropy_bump}. 
For example, as is shown in Fig.~\ref{entropy_bump1}, 
the point $\phi=\pi/3$, $\theta=0,$ and $f=0.8$ appears to be critical, 
but for larger system sizes is shown to be gapped.
We base this conclusion on the appearance of a saturation plateau in the profile of the EE scaling as a function of subsystem size.
As a comparative check, we went deeper into the critical regime 
(i.e. $f=0.9$). As can be seen in Fig.\ref{entropy_bump2} 
and as is expected, we found no such plateau in the EE.


\section{Lifshitz Transition for free fermions}\label{app:lifshitz}
For comparison with our discussion of the Lifshitz transition in the chiral clock model we consider a version with 1D free fermions hopping on a chain with nearest neighbor and next nearest neighbor hopping. As the n.n.n. hopping is increased additional Fermi-points can enter the spectrum and eventually hit the chemical potential which leads to a Lifshitz transition of the Fermi-surface topology.
As our model we consider free fermions with next nearest neighbor hopping. 
\begin{equation}
H = - \sum_n  \left[c^{\dagger}_{n+1} c_n + t  c^{\dagger}_{n+2} c_n + h.c.\right]
\end{equation}
Here, $t$ is the parameter for the next nearest hopping. The energy spectrum of this model is $E=-2\cos(k)-2t\cos(2k)$. When $t$ increases from zero, the topology of the Fermi surface at zero energy changes from two points to four points at $t=1$, which is the Lifshitz transition.

\begin{figure}[htpb]
\centering
\includegraphics[width=\linewidth]{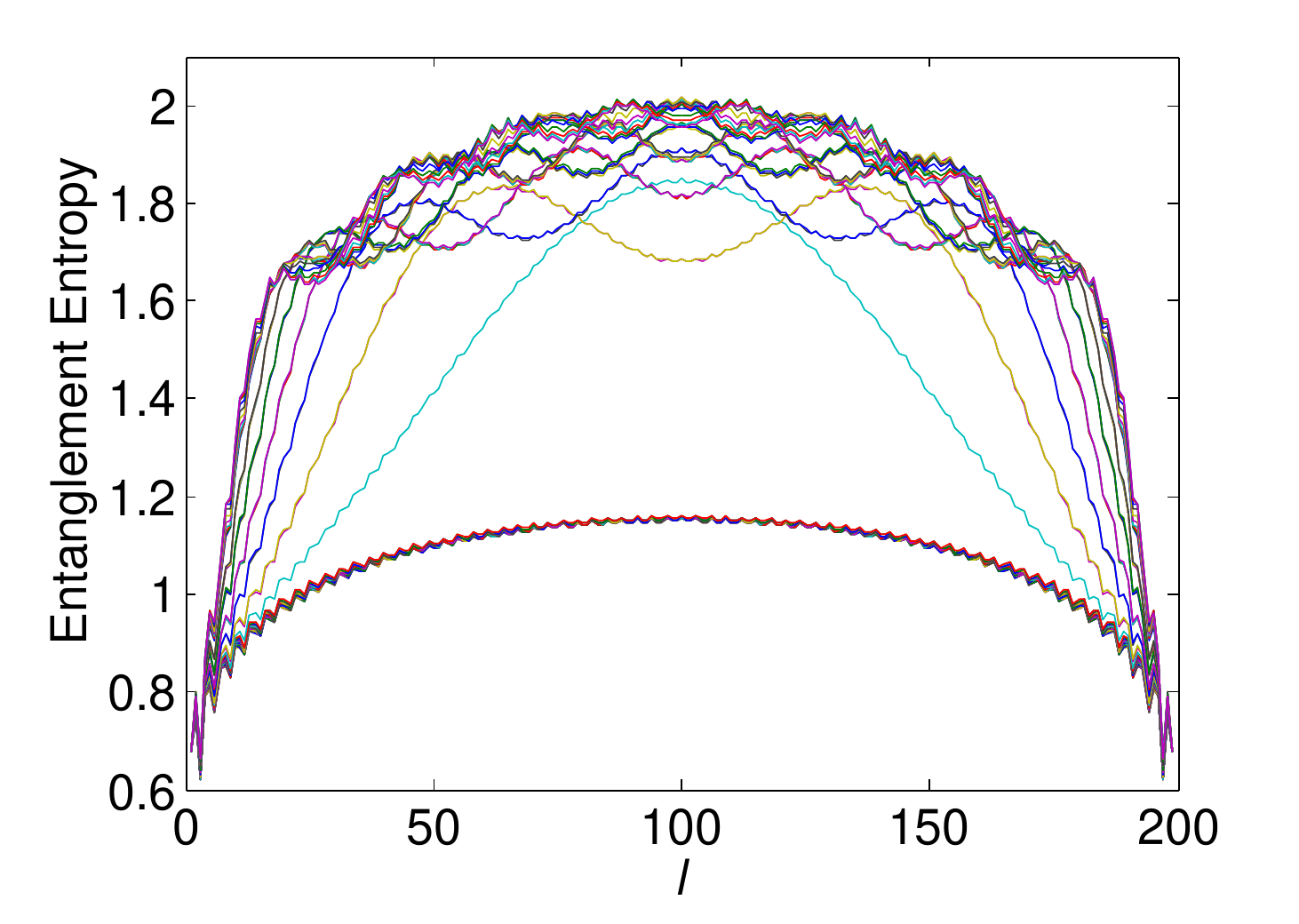}
\caption{Entanglement entropy as a function of block size for different next nearest hopping $t$ with the open boundary condition. $t$ is chosen from 0.97 to 1.03 with a step of 0.001. The lower part corresponds to $t\le1$. The cyan peak appears in the middle is of $t=1.001$.} 
\label{lifshitzopen}
\end{figure}

\begin{figure}[htpb]
\centering
\includegraphics[width=\linewidth]{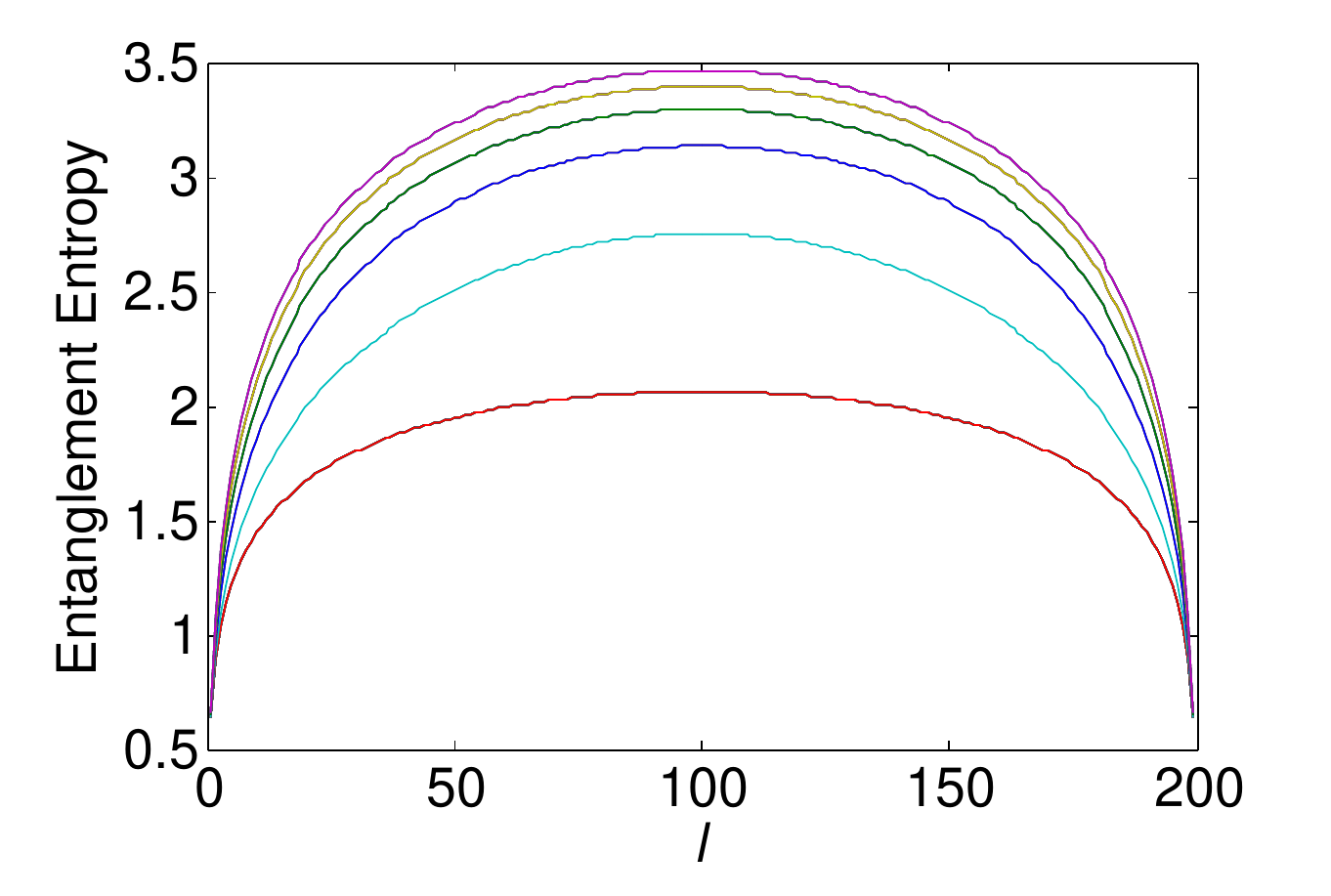}
\caption{Entanglement entropy as a function of block size for different next nearest hopping $t$ with the periodic boundary condition. $t$ is chosen from 0.97 to 1.03 with a step of 0.001. For $t\le1$, all the curves collapse into the lowest curve in the figure. The entanglement entropy increases as we increase $t$. Clearly, there are some steps for such increase.} 
\label{lifshitzperiodic}
\end{figure}

We calculate the entanglement entropy of this model with open boundaries and the periodic boundaries at half filling. The results are shown in Fig.~\ref{lifshitzopen} and Fig.~\ref{lifshitzperiodic} and one can immediately recognize the pattern of oscillations that we saw earlier for the chiral clock model. One interesting thing to notice is that the oscillations go away when we use periodic boundary conditions. 
 This model, and the related entanglement properties, are carefully studied in Ref. \onlinecite{rodney2013}. For periodic boundary conditions the curves gradually increase from the scaling form with $c=1$ to a scaling form of $c=2$ which is the result expected for two sets of left and right movers at the Fermi-level.

\bibliography{refs}{}

\end{document}